\documentclass[11pt,a4paper]{scrartcl}

\usepackage{CLICdp}

\usepackage{CLICdp_definitions}
\usepackage{csquotes}
\usepackage{hyperref}
\usepackage{cleveref}
\usepackage{amssymb}
\DeclareUnicodeCharacter{2212}{-}
\newcommand{\sss}{\scriptscriptstyle}
\newcommand{\mW}{m_{\sss W}}
\newcommand{\sW}{s_{\sss W}}
\newcommand{\cW}{c_{\sss W}}

\title{CP-violating Higgs-gauge boson couplings in $H\nu \bar{\nu}$ production at three energy stages of CLIC}

\clicdppub{2019}{004}  

\date{\today}

\addauthor{O. Karadeniz\thanks{ozgunkdz@gmail.com}}{\institute{1}}
\addauthor{A. Senol\thanks{senol\_a@ibu.edu.tr}}{\institute{1}}
\addauthor{K. Y.  Oyulmaz\thanks{kaan.yuksel.oyulmaz@cern.ch}}{\institute{1}}
\addauthor{H. Denizli\thanks{denizli\_h@ibu.edu.tr}}{\institute{1}}

\addinstitute{1}{Department of Physics, Bolu Abant Izzet Baysal University, 14280, Bolu, Turkey}

\abstract{A phenomenological study of CP-violating dimension-six operators via the $e^+e^-\to\nu \bar{\nu} H$ process is performed in a model-independent Standard Model effective field theory framework at all energy stages of CLIC using the updated baseline integrated luminosities.  All signal and relevant background events  are generated in MadGraph and passed through PYTHIA for parton showering and hadronization at parton level. Detector effects are considered via tuned CLIC detector cards in Delphes. Since we reconstruct the Higgs boson from a pair of b-jets, limits on CP-violating dimension-six couplings are obtained at three $b$-tagging working points: tight, medium and loose defined in the CLIC Delphes card for all three energy stages of CLIC. Our best 95~\% C.L. limits at the loose working point (90~\%  b-tagging efficiency) on $\tilde c_{HW}$ and $\tilde c_{HB}$ are $[-7.0\times10^{-3};7.0\times10^{-3}]$ and $[-3.0\times10^{-2};3.0\times10^{-2}]$, respectively at the 3~TeV energy stage of CLIC with an integrated luminosity of 5.0~ab$^{-1}$. Considering a 0.3~\% systematic uncertainty from possible experimental sources worsens the limits on these couplings by a factor of two.

}

\titlecomment{This work was carried out in the framework of the CLICdp Collaboration} 

\begin{document}

\titlepage

\section{Introduction}
Although the observation of a scalar particle with mass about 125 GeV compatible with the Standard Model (SM) Higgs boson at the Large Hadron Collider (LHC) \cite{Aad:2012tfa,Chatrchyan:2012xdj} marked a milestone in particle physics, evidence for new physics beyond the SM has not been observed yet in the analysis of combined  ATLAS and CMS data to probe the couplings of the Higgs boson. Therefore, one of the main topics of the High-Luminosity LHC (HL-LHC) program and envisaged future high-energy collider projects will be the precise measurement of the Higgs-boson couplings to the SM particles.

In the SM framework, the experimental data is currently consistent with a CP-even hypothesis and the charge conjugation-parity (CP) violation is described by Cabibbo-Kobayashi-Maskawa (CKM) matrix \cite{Cabibbo:1963yz,Kobayashi:1973fv} with a single complex phase in the Yukawa sector. However, the origin of the baryon asymmetry of the universe can not be explained by the CP violation in the SM \cite{Dine:2003ax}. An extended Higgs sector together with CP-violation beyond the Standard Model (SM) is one of the conceivable options to explain the baryon asymmetry of the Universe. Especially, the couplings of Higgs to SM gauge bosons and/or fermions are interesting possibilities which contain new sources of CP-violation. A well-known approach of searching for new physics in a model-independent way is the SM Effective Field Theory (EFT).
The basic principle of this approach is that all new physics contributions to the SM are included with higher-dimensional operators conforming to $SU(3)_C\times SU(2)_L\times U(1)_Y$  SM gauge symmetry \cite{Buchmuller:1985jz,Grzadkowski:2010es}. The possibility of CP-violating couplings involving higher-dimensional interaction terms containing the Higgs and gauge boson pairs cannot be discarded in the investigation of new physics. 
Searches for CP-violating Higgs-gauge boson couplings via higher-dimensional operators were previously performed in many rewarding studies at experimental LHC data \cite{Aaboud:2018xdt,Aad:2015tna,Sirunyan:2019nbs,Sirunyan:2019twz} and  phenomenologically at LHC \cite{Godbole:2007cn,Christensen:2010pf,Desai:2011yj,Godbole:2013lna,Brod:2013cka,Freitas:2012kw,Gavela:2014vra,Dolan:2014upa,Dwivedi:2015nta,Rong:2016lmo,Bernlochner:2018opw,Cirigliano:2019vfc}, at HE-LHC and HL-LHC \cite{Ferreira:2016jea,Dwivedi:2016xwm,Cepeda:2019klc}, at future $e^+e^-$ 
\cite{Jones:1979bq,Han:2000mi,Hagiwara:2000tk,Ginzburg:2000rm,Biswal:2005fh,Rao:2006hn,Dutta:2008bh,Biswal:2008tg,Sahin:2008jc,Biswal:2009ar,Rindani:2009pb,Beneke:2014sba,Amar:2014fpa,Craig:2015wwr,Kumar:2015eea,Sahin:2017mha} and $ep$ colliders \cite{Biswal:2012mp}.

The precision measurements of the Higgs boson couplings with the other SM particles at the LHC and planned future colliders will give us detailed information about its true nature. Among the proposed future collider projects, the Compact Linear Collider (CLIC) is a mature option for a future linear electron-positron collider \cite{Linssen:2012hp,Charles:2018vfv}, which is currently under development as a possible large-scale installation at CERN.  Additionally, CLIC comes to the fore with features such as i) allowing the precise determination of the properties of the Higgs boson well beyond the precision of the HL-LHC \cite{Abramowicz:2016zbo,Robson:2018zje}, ii) being a unique and innovative two-beam acceleration technique that can reach accelerating gradients of 100~MV/m, iii) providing high-luminosity $e^+e^-$ collisions at a series of center-of-mass energy stages from a few hundred GeV up to 3~TeV, iv) benefiting from the clean experimental environment and good knowledge of the initial state to allow precise measurements of many reactions. 
To diversify the physics opportunities, CLIC will be operated in several center-of-mass energy stages. A first stage at 380~GeV center of mass energy gives a suitable platform not only to the Higgsstrahlung process in which $e^+e^-$ collisions enable a unique Higgs physics programme, but also to perform  a scan over the $t\bar t$ production threshold \cite{Roloff:2018dqu, Abramowicz:2018rjq}. The higher-energy stages, currently assumed to be at 1.5~TeV (second stage) and 3~TeV (third stage), provide a unique sensitivity for a large number of new physics scenarios. 
Updated integrated luminosities for these three energy stages,  based on accelerator ramp-up and up-time scenarios, are 1.0~ab$^{-1}$, 2.5~ab$^{-1}$ and 5.0~ab$^{-1}$, respectively \cite{Robson:2018zje}. Each stage is planned to run seven or eight years with a two-year commissioning between  so that the physics program will be completed  within approximately 25-30 years.
  
Since dimension-6 CP-even operators have been studied in CLIC \cite{Charles:2018vfv,Denizli:2017pyu}, we investigate the effect of CP-violating dimension-6 operators of the $HWW$, $H\gamma\gamma$ and $HZZ$ vertices defined by an SM EFT Lagrangian in the $e^+e^-\to \nu \bar{\nu} H$ production process at the three energy stages of CLIC in this study. The rest of the paper is organized as follows: in the next section we give details of the Effective Field Theory approach and operators of the dimension-6 CP-violating interactions of the Higgs boson and electroweak gauge boson in the Strongly-Interacting Light Higgs (SILH) basis. In section 3, the event selection criteria and cut optimization of the signal and relevant background processes are discussed for each stage of CLIC. The sensitivity estimations of CP-violating dimension-6 Higgs-Gauge boson couplings are given in section 4. Finally, we conclude and compare our obtained limits with current experimental results in section 5.
\section{Effective Operators}
It is well known that all operators which define quark and lepton fields along with a single Higgs doublet field interacting via an $SU(3)_C\times SU(2)_L\times U(1)_Y$ SM gauge symmetry are restricted to have mass dimension of four or less in the SM Lagrangian ( $\mathcal{L}_{\textrm SM}$ ). In the Effective Field Theory (EFT) approach, the SM Lagrangian is extended with higher-dimensional operators having coefficients of inverse powers of mass ($\Lambda$), and hence are suppressed if this mass is large compared to the reachable experimentally energies;
\begin{eqnarray}
\mathcal{L} = \mathcal{L}_{\textrm SM} + \sum_{i}\frac{c_{i}}{\Lambda^2}\mathcal{O}_{i}
\end{eqnarray}
where $\Lambda$ is  the new physics scale, $\mathcal{O}_{i}$ are the dimension-six operators, and the coefficients $c_i$ are dimensionless parameters of the new physics coupling to the SM particles. It is also noted that we ignore the dimension-5 operators responsible for generating Majorana neutrinos and are only concerned with the extended Lagrangian with dimension-6 operators.
The most general gauge-invariant dimension-6 Lagrangian $\mathcal{L}$ can be expressed in a convenient basis of independent operators $\mathcal{O}_{i}$ using  normalized Wilson coefficients as $\bar c_{i}=c_{i}/\Lambda^2$ that are free parameters \cite{Giudice:2007fh,Contino:2013kra,Alloul:2013naa}. In this work, we consider the dimension-6 CP-violating interactions of the Higgs boson and electroweak gauge boson in the SILH basis as \cite{Alloul:2013naa}:
\begin{eqnarray}\label{CPodd}
  {\cal L}_{CPV} &= & \frac{i g\ \tilde c_{HW}}{m_W^2}  D^\mu \Phi^\dag T_{2k} D^\nu \Phi {\widetilde W}_{\mu \nu}^k
  + \frac{i g'\ \tilde c_{ HB}}{m_W^2} D^\mu \Phi^\dag D^\nu \Phi {\widetilde B}_{\mu \nu}
  + \frac{g'^2\  \tilde c_{\gamma}}{m_W^2} \Phi^\dag \Phi B_{\mu\nu} {\widetilde B}^{\mu\nu} \nonumber\\
  &+&\frac{g_s^2\ \tilde c_{ g}}{m_W^2}      \Phi^\dag \Phi G_{\mu\nu}^a {\widetilde G}^{\mu\nu}_a
  +\frac{g^3\ \tilde c_{ 3W}}{m_W^2} \epsilon_{ijk} W_{\mu\nu}^i W^\nu{}^j_\rho {\widetilde W}^{\rho\mu k}
  +\frac{g_s^3\ \tilde c_{ 3G}}{m_W^2} f_{abc} G_{\mu\nu}^a G^\nu{}^b_\rho {\widetilde G}^{\rho\mu c} \ ,
\end{eqnarray}
where the dual field strength tensors are defined by
\begin{eqnarray}
  \widetilde B_{\mu\nu} = \frac12 \epsilon_{\mu\nu\rho\sigma} B^{\rho\sigma} \ , \quad
  \widetilde W_{\mu\nu}^k = \frac12 \epsilon_{\mu\nu\rho\sigma} W^{\rho\sigma k} \ , \quad
  \widetilde G_{\mu\nu}^a = \frac12 \epsilon_{\mu\nu\rho\sigma} G^{\rho\sigma a} \ .
\end{eqnarray}
and $\Phi$ is the Higgs field containing a single $SU(2)_L$ doublet of fields; $B^{\mu\nu}=\partial_\mu B_\nu-\partial_\nu B_\mu$  and $W^{\mu \nu}=\partial_\mu W_\nu^k-\partial_\nu W_\mu^k+g\epsilon_{ijk}W_\mu^iW_\nu^j$ are the electroweak field strength tensor and $G^{\mu\nu}$ is the strong field strength tensors; $g'$, $g$ and $g_s$ denote coupling constant of  $U(1)_Y$, $SU(2)_L$ and $SU(3)_C$ gauge fields, respectively; the generators of $SU(2)_L$ in the fundamental representation are given by $T_{2k}=\sigma_k/2$ and $\sigma_k$ are the Pauli matrices. 
The SM EFT Lagrangian (Eq.(\ref{CPodd})) containing the Wilson coefficients in the SILH basis of dimension-6 CP-violating operators can be defined in terms of the mass eigenstates after electroweak symmetry breaking (Higgs boson, W, Z, photon, etc.) as follows:
\begin{eqnarray}\label{gbase}
  {\cal L}_{CPV}& = &\ 
    - \frac{1}{4} \tilde g_{\sss hgg} G^a_{\mu\nu} \tilde G^{\mu\nu} h
    - \frac{1}{4} \tilde g_{\sss h\gamma\gamma} F_{\mu\nu} \tilde F^{\mu\nu} h
    - \frac{1}{4} \tilde g_{\sss hzz} Z_{\mu\nu} \tilde Z^{\mu\nu} h \nonumber\\
    &-& \frac{1}{2} \tilde g_{\sss h\gamma z} Z_{\mu\nu} \tilde F^{\mu\nu} h
    - \frac{1}{2} \tilde g_{\sss hww} W^{\mu\nu} \tilde W^\dag_{\mu\nu} h,
\end{eqnarray}
where $W_{\mu\nu}$, $Z_{\mu\nu}$ and $F_{\mu\nu}$ are the field strength tensors of $W$-boson, $Z$-boson and photon, respectively. The effective couplings in gauge basis defined as dimension-6 operators are given in Table~\ref{mtable}.
\begin{table}[h]
\caption{The relations between Lagrangian  parameters in the mass basis (Eq.\ref{CPodd}) and in the gauge basis (Eq. \ref{gbase}). ($c_W\equiv\cos \theta_W$, $s_W\equiv\sin \theta_W$)}  
\begin{center}
\label{mtable}
\begin{tabular}{lcl}\hline\hline
$\tilde g_{\sss hgg}$= $-\frac{4 \tilde c_{\sss g} g_s^2 v}{\mW^2}$  && $\tilde g_{ h\gamma\gamma}$  
     $= -\frac{8 g \tilde c_{ \gamma} \sW^2}{\mW}$ \\
 $\tilde g_{ hzz}$ =$\frac{2 g}{\cW^2 \mW} \Big[ \tilde c_{ HB} \sW^2 - 4 \tilde c_{ \gamma} \sW^4 + \cW^2 \tilde c_{ HW}\Big]$&&$\tilde g_{ h\gamma z}$ =$\frac{g \sW}{\cW \mW} \Big[  \tilde c_{HW} - \tilde c_{ HB} + 8 \tilde c_{ \gamma} \sW^2\Big]$    \\
$\tilde g_{\sss hww}$=$\frac{2 g}{\mW} \tilde c_{\sss HW}$&& \\ \hline\hline
     \end{tabular}
\end{center}
\end{table}

\begin{figure}[hbt!]
\includegraphics[scale=0.41]{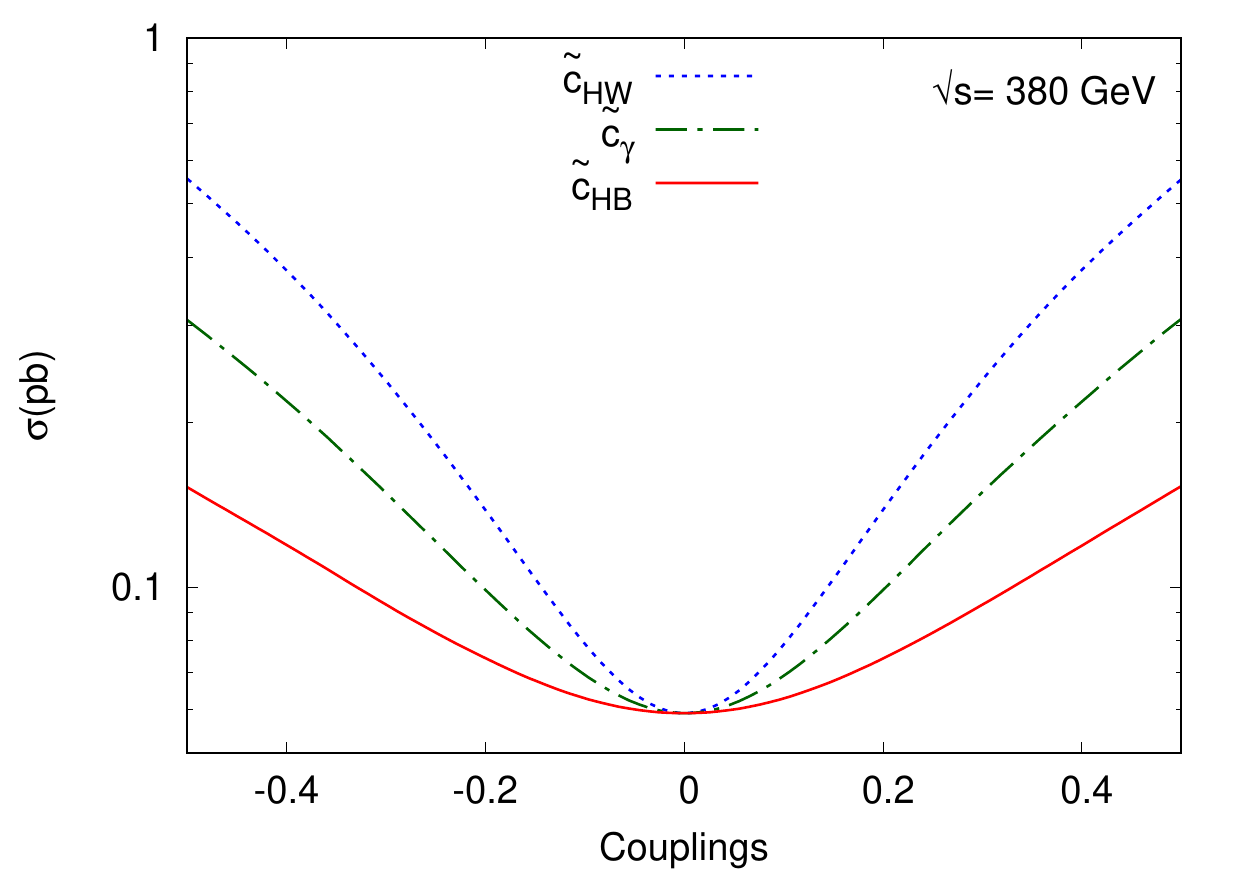} \includegraphics[scale=0.41]{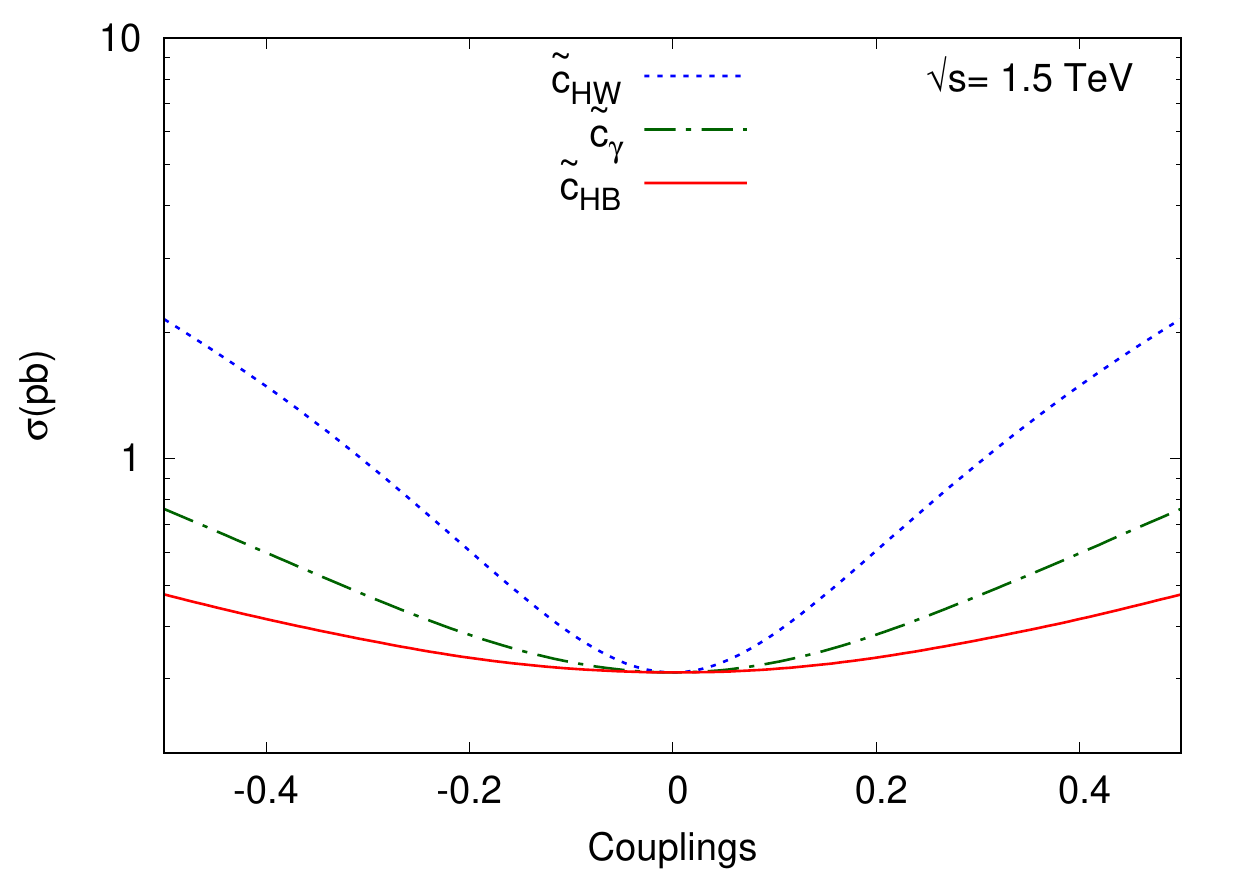} \includegraphics[scale=0.41]{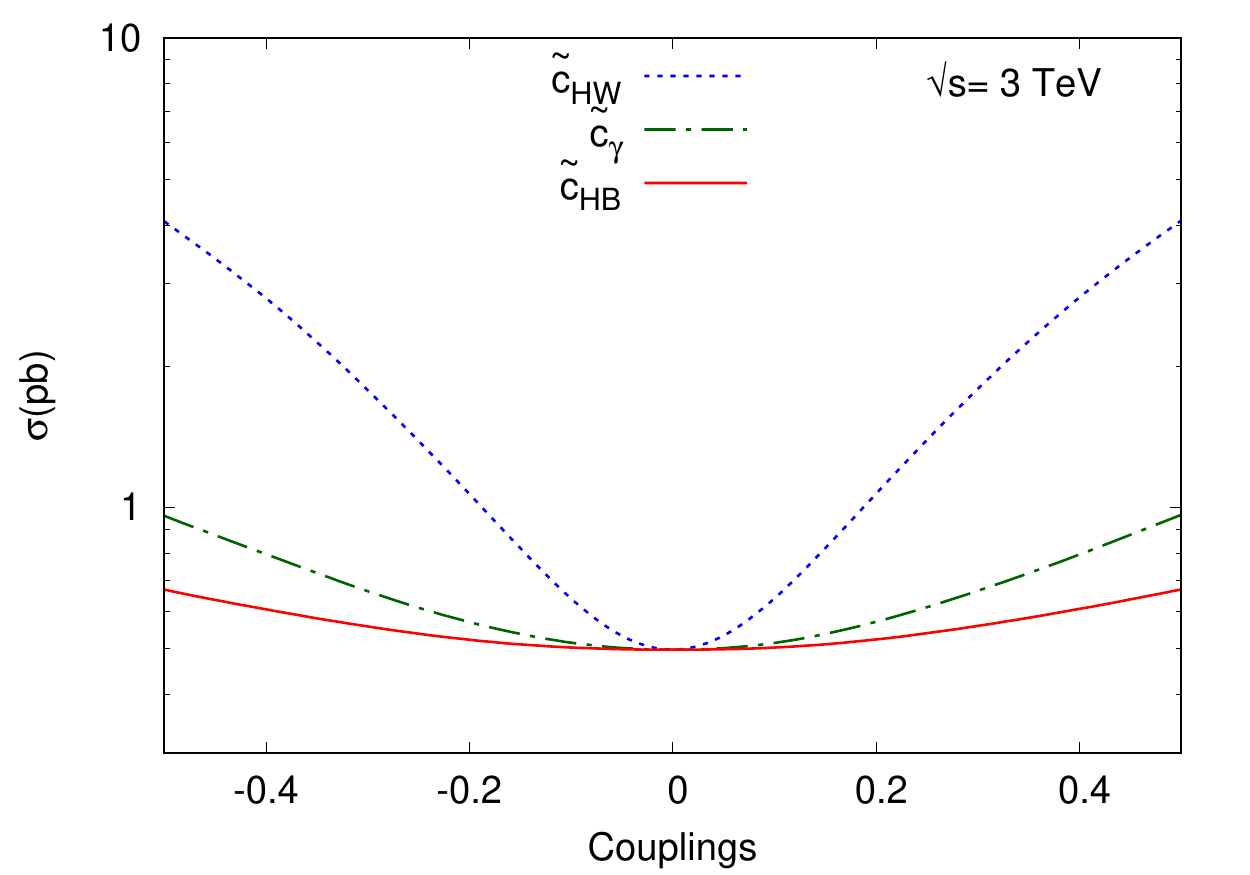} 
\caption{ The total cross section as a function of CP-violating $\tilde{c}_{HW}$, $\tilde{c}_{HB}$ and $\tilde{c}_{\gamma}$ couplings for the $e^+e^-\to \nu \bar{\nu} H$ process at the CLIC with $\sqrt s$=380~GeV, $\sqrt s$=1.5~TeV and $\sqrt s$=3~TeV. \label{fig1}}
\end{figure}

The parametrization of Ref.\cite{Alloul:2013naa} which is based on the formulation given in Ref.\cite{Contino:2013kra} is considered in our analysis. Since it chooses to remove two fermionic invariants while retaining all the bosonic operators, the parametrization is not complete as explained in Ref. \cite{Alonso:2013hga, Brivio:2017bnu}.  
The main purpose of this paper is to estimate the direct sensitivity to $\tilde c_{HW}$, $\tilde c_{HB}$ and $\tilde c_{\gamma}$ couplings without considering higher-order electroweak effects. For this purpose, the effects of the dimension-6 CP-violating operators on $H \nu \bar{\nu} $ production mechanism in $e^+e^-$ collisions are investigated using the Monte Carlo simulations with leading order in \verb|MadGraph5_aMC_v2.6.3.2@NLO| \cite{Alwall:2014hca}.  The described CP-violating operators in the effective Lagrangian of Eq.(\ref{CPodd}) are implemented into the \verb|MadGraph5_aMC@NLO| based on FeynRules \cite{Alloul:2013bka} and the UFO \cite{Degrande:2011ua} framework.
The cross sections of $e^+e^-\to\nu \bar{\nu} H$ process at generator level as a function of  $\tilde c_{HW}$, $\tilde c_{HB}$ and $\tilde c_{\gamma}$ couplings for three center of mass energy stages of CLIC; 380~GeV, 1.5 and 3~TeV are given in Fig.\ref{fig1}. The quoted cross sections do not include the effects of initial state radiation (ISR) and beamstrahlung. In this figure, we vary dimension-6 CP-violating operators individually and calculate the contributions to the corrections from new physics. Since one coefficient at a time is varied in the calculation of cross section, only quartic contributions are taken into account. It  can be easily seen that the contribution of the $\tilde c_{HW}$ coupling to the SM increases with center of mass energy even in a small value region for the $e^+e^-\to\nu \bar{\nu} H$ process.
 \section{Signal and Background Analysis}
 In this section, we investigate the sensitivity of CLIC at three center of mass energy stages for CP-violating $\tilde c_{HW}$, $\tilde c_{HB}$, $\tilde c_{\gamma}$ effective couplings through the $e^+e^- \rightarrow \nu\bar{\nu}H$ process and relevant backgrounds assuming Higgs boson decays to a pair of $b$-quarks. The effective CP-violating dimension-6 couplings and SM contribution ($S+B_H$) in the $e^+e^-\to\nu \bar{\nu} H$ process are taken into account. The values of CP-violating dimension-6 couplings considered in this study are very small 
so the total decay width under influence of these operators are close to the SM expectations. Therefore, the influence of CP-violating dimension-6 couplings on the branching ratio of $H\to b\bar b$ is neglected in the analysis. The following relevant backgrounds are included in the analysis. \textbf{i)} The same final state as the considered signal process including only SM contribution is the $e^+e^-\to\nu \bar{\nu} H$ process, which is labelled $B_H$. \textbf{ii)} The production of two Z bosons is labeled as $B_{ZZ}$, considering one $Z$ decaying to $b\bar b$ while the other decays to $\nu\bar \nu$. \textbf{iii)} The $W$ boson pair production is labeled as $B_{WW}$, considering one $W$ decaying to $b\bar b$ while the other decays to $l\nu$. 
 \begin{figure}[hbt!]
\includegraphics[scale=0.22]{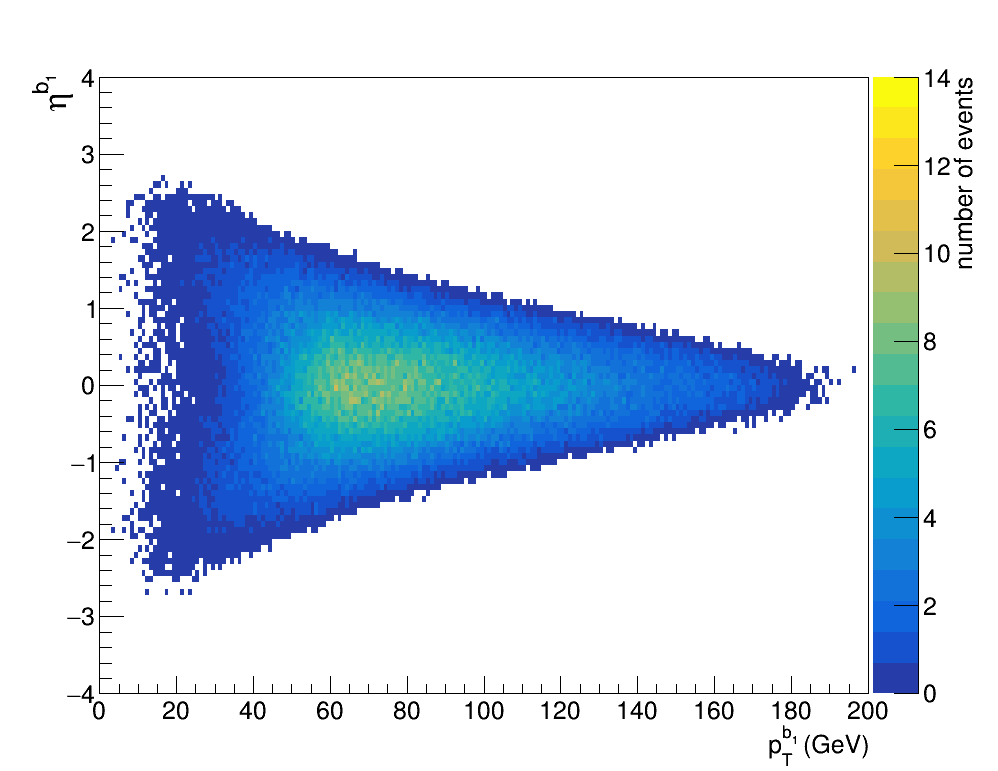} 
\includegraphics[scale=0.22]{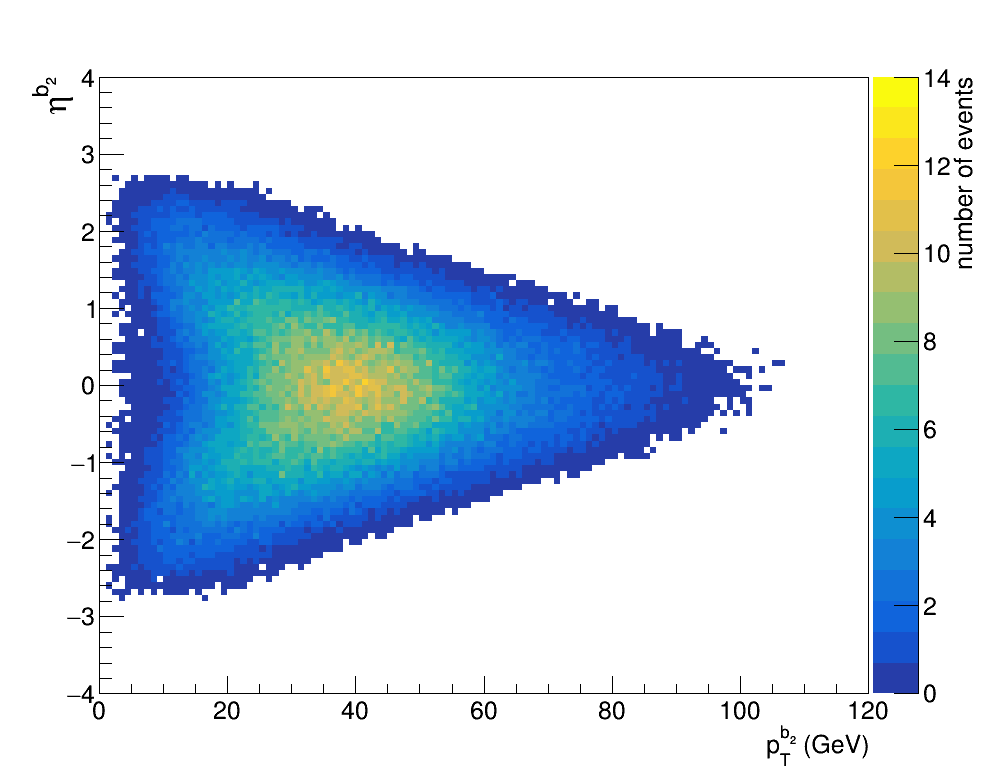} 
\caption{ The pseudo-rapidity versus transverse momentum distribution of leading ($b1$) and sub-leading ($b2$) b-tagged jets for SM $e^+e^-\to \nu \bar{\nu} H$ process with defined WP (90~\% b-tagging efficiency) in CLIC Delphes card at $\sqrt s$=380~GeV \label{fig2a}}
\end{figure} 
\textbf{iv)} $B_{tt}$ is the pair production of the top quark process i.e. , $e^+e^-\to t \bar t$ in which one of the top quark  (anti-top quark) decays to $W^+b (W^{-} \bar b)$, where the leptonic decay channel of $W^{\pm}$ is considered. \textbf{v)} The hadronic decay channel of the $Z$ boson in the $e^+e^-\to\nu \bar{\nu} Z$ process is taken into account and labelled $B_{Z\nu\nu}$. As shown in Ref \cite{Abramowicz:2016zbo}, one can expect to see significant contribution to the background due to $e\gamma$ and $\gamma\gamma$ collisions. In our analysis framework, we generate events via Madgraph which does not include photons from Beamstrahlung. Therefore, we neglect these backgrounds in our analysis. The cross section of the considered backgrounds in our analysis are given in Table \ref{tab5}.
\begin{table}
\begin{center}
\caption{The cross sections of the relevant backgrounds in pb before applying any cuts \label{tab5}}
\begin{tabular}{lccccc}\hline\hline
$\sqrt s$  &$B_H$ &$B_{ZZ}$&$B_{WW}$&$B_{tt}$&$B_{Z\nu\nu}$  \\ \hline
   380 GeV      & 0.059&0.597&10.229&0.618&0.419\\
   1.5    TeV   & 0.310&0.079&1.437&0.079&1.339\\
       3 TeV   & 0.497&0.026&0.47&0.020&2.16 \\\hline\hline
\end{tabular}
\end{center}
\end{table}
All signal and background events (500k for each) are generated in \verb|MadGraph5_aMC@NLO|  and passed through PYTHIA 8.2 for parton showering, hadronization and decay of unstable particles \cite{Sjostrand:2014zea}. We use the Delphes 3.4.1 \cite{deFavereau:2013fsa}  for a fast simulation of detector response with tuned CLIC detector cards \cite{Leogrande:2019qbe, Tehrani:2017}. There are three cards, designed for each center-of-mass energy stage of CLIC: $\sqrt s=$380~GeV, 1.5~TeV and 3~TeV. Some properties of the cards are as follows. Jets are clustered with the Valencia Linear Collider (VLC) algorithm \cite{Boronat:2014hva,Boronat:2016tgd} in exclusive mode with a fixed number of jet ($N=2,3,4,5,6$ where $N$ corresponds to the number of partons expected in the tree level final state) and five different cone size parameters ($R=0.5,0.7,1.0,1.2,1.5$) with $\gamma$=$\beta$=1 using FastJet \cite{Cacciari:2011ma}. The b-tagging efficiency and misidentification rates implemented in these cards are discussed in Ref. \cite{Tehrani:2015tla,AlipourTehrani:1742993} where the three working points (WP) are defined; the tight WP (50~\% b-tagging efficiency), medium WP (70~\% b-tagging efficiency), and loose WP (90~\% b-tagging efficiency). Misidentification rates for three working points are given as a function of energy and pseudorapidity. For example; In a bin where $E \geqslant 500$ GeV and $1.53 < |\eta| \leqslant 2.09$, misidentification rates are $3\times10^{-3}$, $9\times10^{-3}$ and $5\times10^{-2}$ for the tight, the medium and the loose WP, respectively. In our analysis, we picked $N_{jets}$=2 and $R$=1.0 for three energy stages with the three $b$-tagging working points, tight, medium and loose. Then, all events are analyzed by using the ExRootAnalysis utility \cite{exroot} with ROOT 6.16. \cite{Brun:1997pa}. 
\begin{figure}[hbt!]
\includegraphics[scale=0.22]{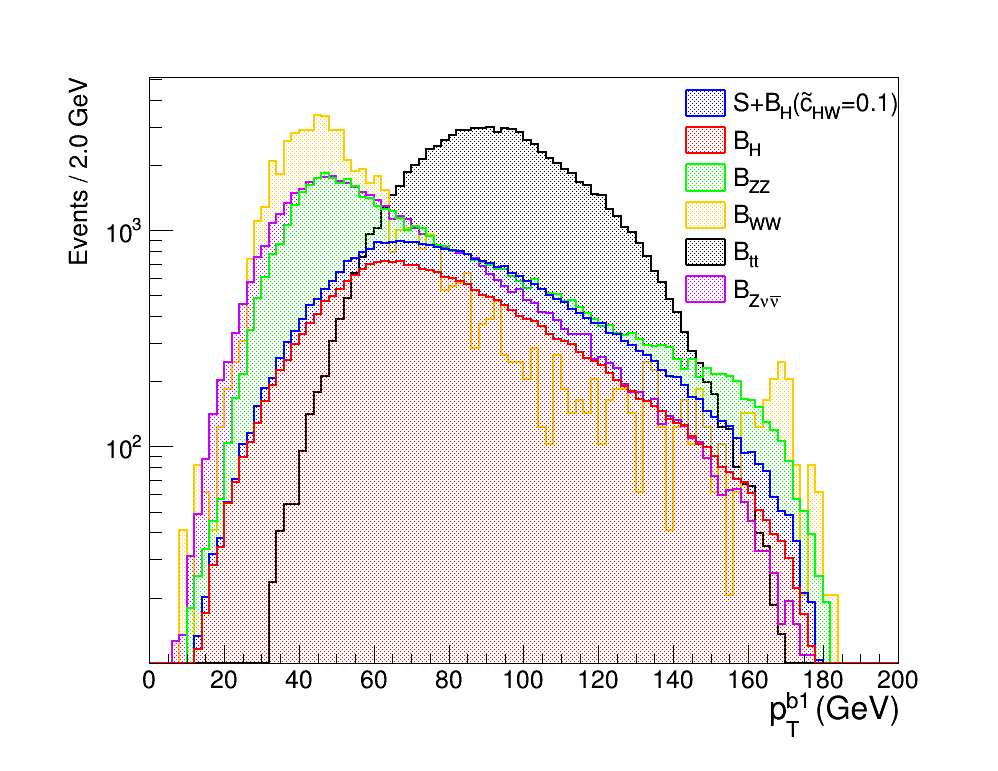} \includegraphics[scale=0.22]{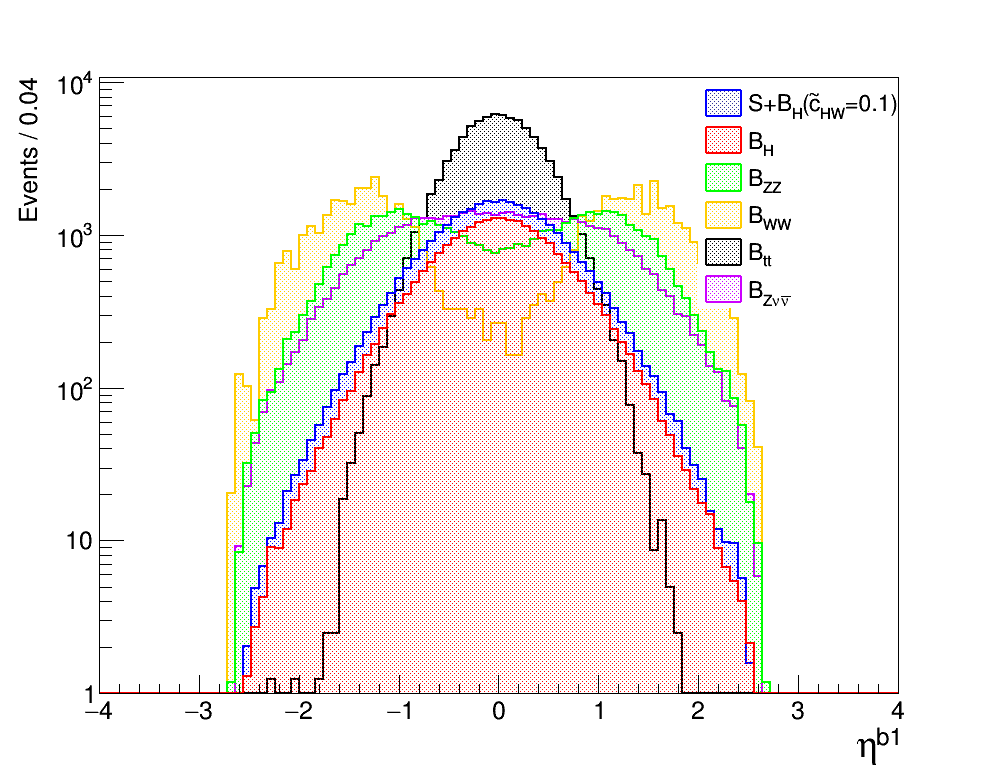}\\ 
\includegraphics[scale=0.22]{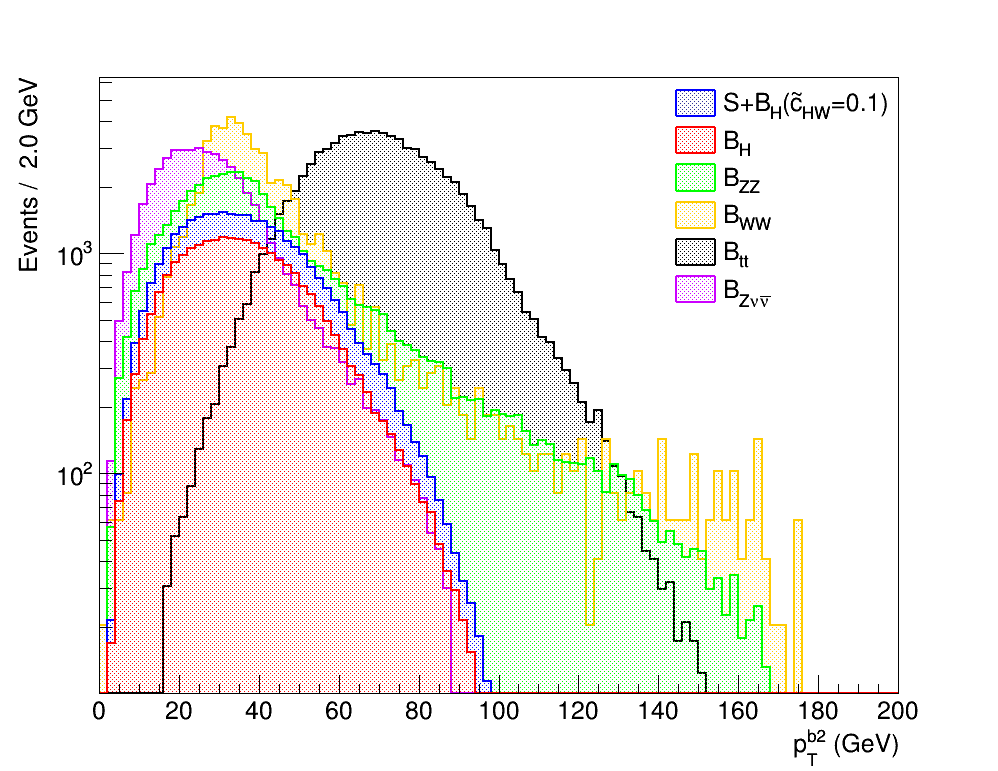} \includegraphics[scale=0.22]{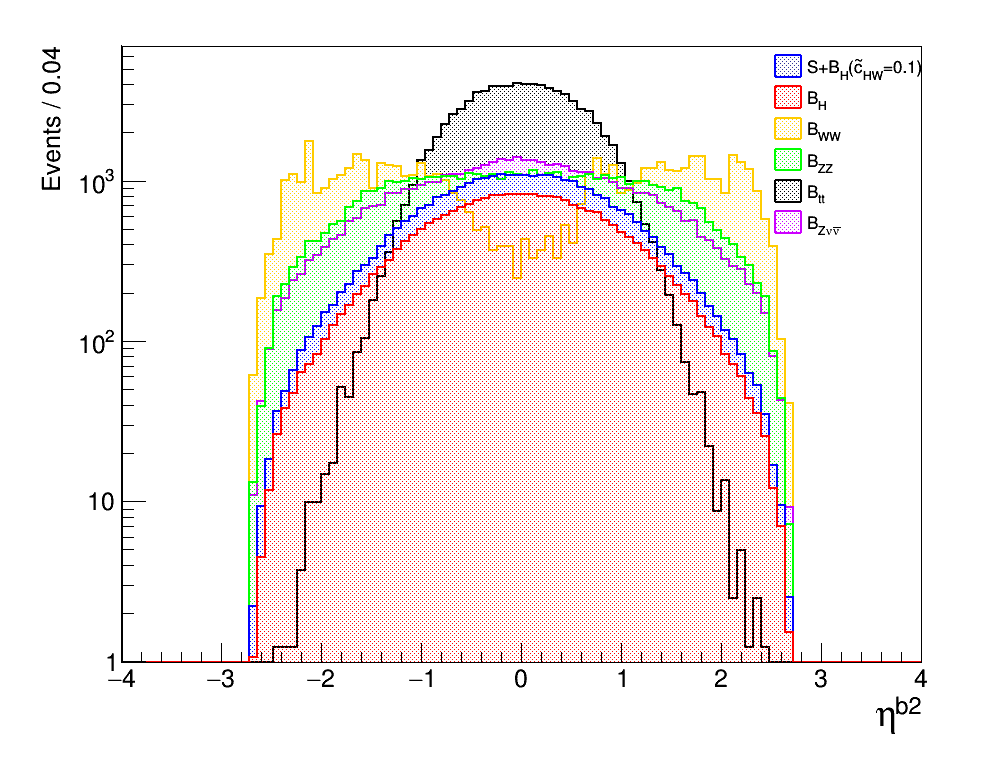} 
\caption{ Normalized distributions of transverse momentum and pseudo-rapidity of tagged b-jets at $\sqrt s=$ 380~GeV; $b1$ (first row) and $b2$ (second row) for signal with $\tilde{c}_{HW} $=0.1 and relevant background processes. \label{fig2}}
\end{figure}
\subsection*{$\sqrt{s}$ = 380~GeV}
In addition to initial jet clustering (i.e, $N_{jets}$=2 and $R$=1.0), events having no charged leptons are selected for further analysis (Cut-0). In order to separate signal and background events we use the following kinematical cuts: \textbf{i)} In exclusive mode, we have two jets which are obtained from subsequent decay of Higgs boson, tagged as $b$-tagged jets. The $b$-tagged jet with the highest transverse momentum ($p_T$) is labeled as $b1$ and the one with lower $p_T$ as $b2$ (Cut-1). The phase space of b-tag jets for the SM background process with the same final state as signal at b-tagging efficiency working points (90~\%) defined in the CLIC Delphes card are shown in Fig.\ref{fig2a}. The transverse momentum and pseudo-rapidity of $b1$ and $b2$ for signal (for $\tilde{c}_{HW} $=0.1 benchmark point) and all relevant background processes taking the loose $b$-tagging working point are shown in Fig.\ref{fig2}. \textbf{ii)}  We select a region in phase space where the transverse momentum of $b1$ is $p_T^{b1}>50$~GeV and $b2$ has $p_T^{b2}>30$~GeV, and the pseudo-rapidity of the $b$-tagged jets is $|\eta^{b1,b2}|$ $\leq 2.0$.
 \begin{figure}[hbt!]
\includegraphics[scale=0.22]{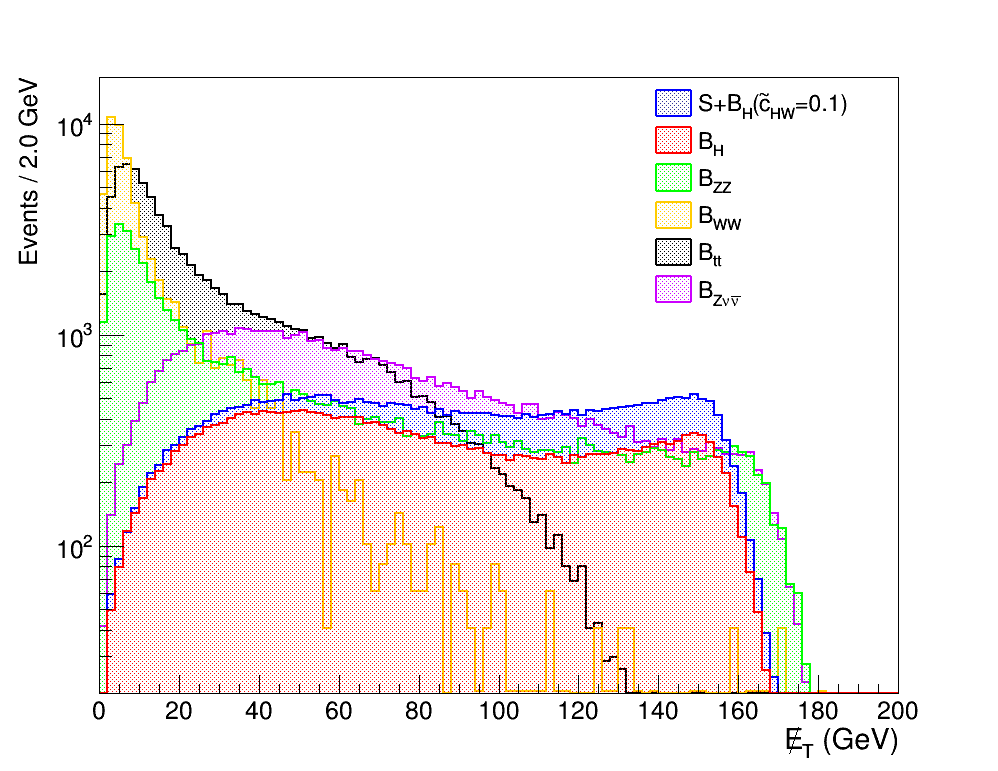} \includegraphics[scale=0.22]{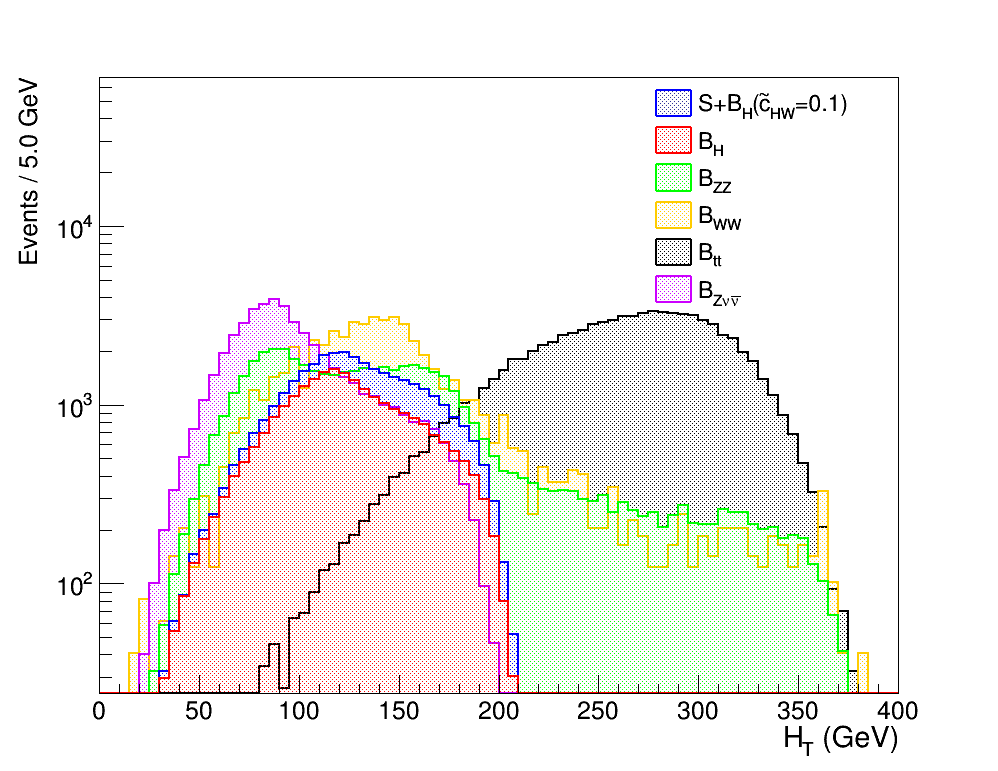}\\  
\includegraphics[scale=0.22] {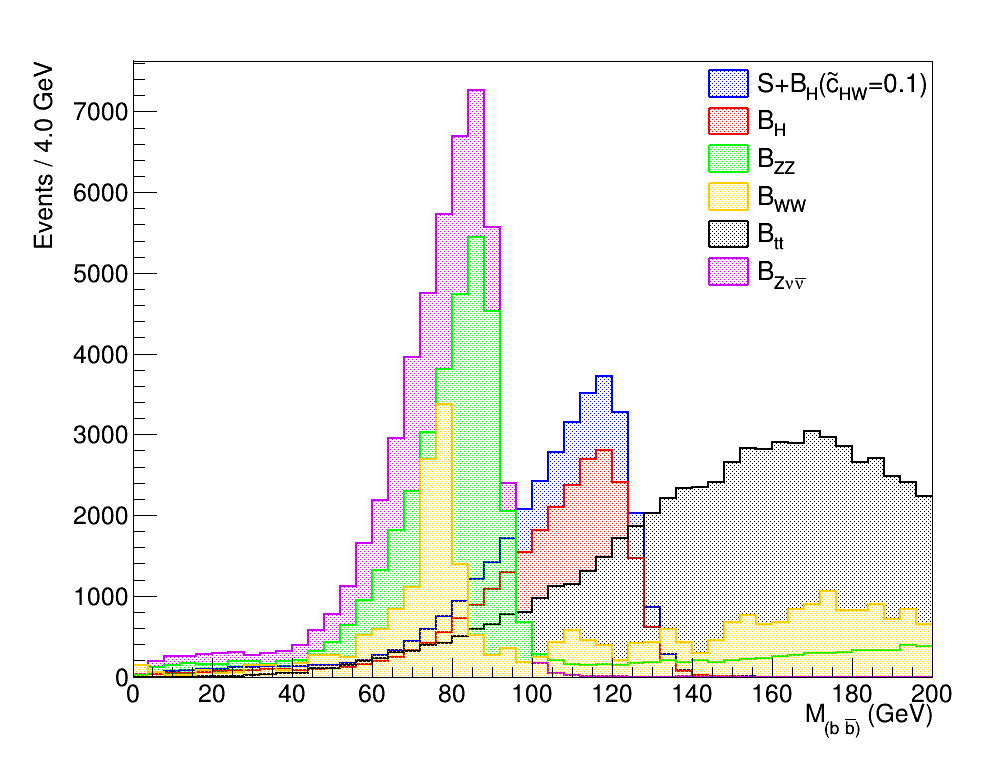}\includegraphics[scale=0.22] {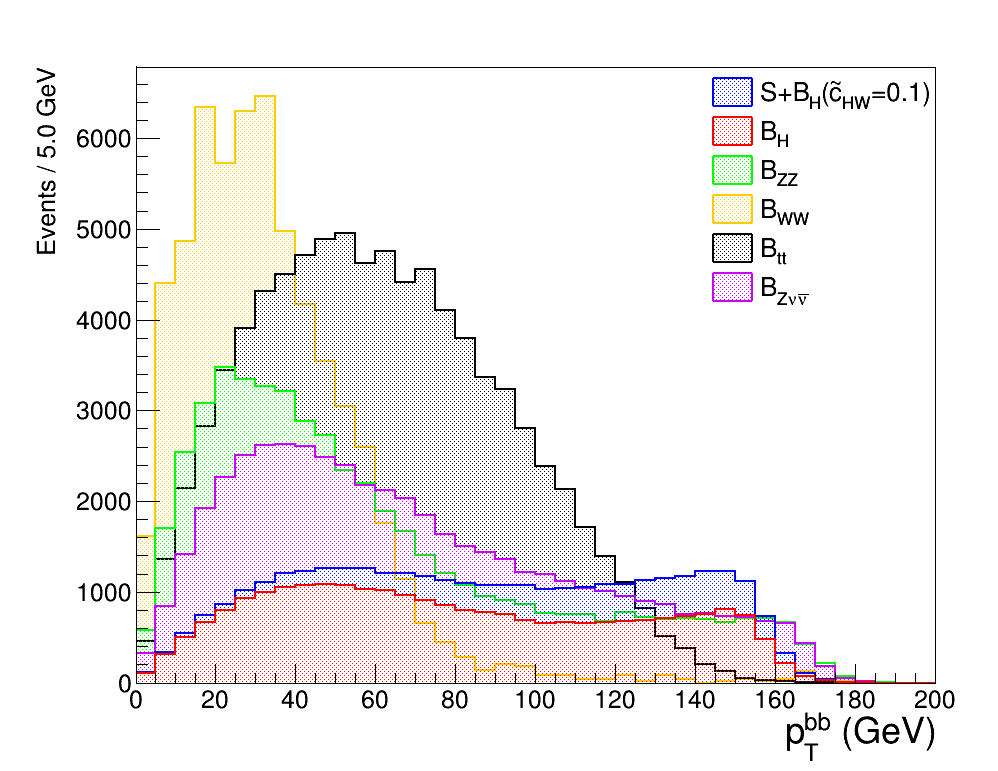} 
\caption{ Normalized distributions of Missing Energy Transverse  (${\not}E_T$), scalar sum of the transverse energy ($H_T$) (in the first row), and invariant mass and transverse momentum of reconstructed Higgs boson at $\sqrt s=$ 380~GeV for signal with $\tilde{c}_{HW} $=0.1 and relevant background processes. \label{fig3}}
\end{figure}
This cut suppresses $B_{ZZ}$ and $B_{Z\nu\bar{\nu}}$ backgrounds. For signal and background processes, distributions of the missing energy transverse (${\not\mathrel{E}}_T$), scalar sum of the transverse energy ($H_T$), the invariant mass and the transverse momentum of the reconstructed Higgs-boson from two $b$-jets are depicted in Fig.\ref{fig3}. Subsequent cuts can be determine from these figures: The missing energy transverse ${\not\mathrel{E}} _T > 30$~GeV  provides a way of reducing the $B_{ZZ}$ and $B_{tt}$ backgrounds at the region with low missing energy transverse (Cut-2). \textbf{iii)} Requiring scalar sum of the transverse energy $(H_T)$  to be  100~GeV $< H_T <$ 200~GeV drastically reduces $B_{tt}$ background (Cut-3). \textbf{iv)} The invariant mass of the reconstructed Higgs-boson from two $b$-jets is required to be 96~GeV $< M_{(b\bar b)} <$ 136~GeV (Cut-4). \textbf{v)} Finally, the transverse momentum of reconstructed Higgs-boson from two $b$-jets $p_T^{bb}>75$ GeV is used to obtain limits on the $\tilde{c}_{HW}$,$\tilde{c}_{HB}$ and $\tilde{c}_{\gamma}$ couplings (Cut-5). The selection criteria and cut flows are summarized in Table \ref{tab2}. The numbers of events after each cut are shown in Table \ref{tab3} for three working points of b-tagging efficiency. As seen from this table, $B_{ZZ}$, $B_{tt}$, $B_{Z\nu\nu}$ backgrounds are reduced more than signal $(S + B_H)$ and background $B_H$.
\begin{table}
\caption{Event selection criteria and applied kinematic cuts used for the analysis at three energy stages of CLIC \label{tab2}}
\begin{center}
\begin{tabular}{lcccccc}\hline\hline
Cuts&$\sqrt s=$380~GeV&\vline& $\sqrt s=$1.5~TeV&\vline&$\sqrt s=$3~TeV \\ \hline 
        &                              &&Jet clustering: VLC with $\beta=\gamma=1.0$ R=1.0 &&   \\
Cut-0 &&& Exclusive clustering with $N_j=2$ Jets &&\\
&& & energy scale is assumed to be 1.0 &&\\
& &&lepton vetos &&\\ \hline 
Cut-1  & &&requiring two b-tagged jets &&\\ \hline
Cut-2       &&& $p_{T}^{b1} > 50$~GeV, $p_{T}^{b2} > 30$~GeV &&\\
  &&&  $\eta^{b1,\,b2}\leqslant 2.0$, ${\not}E_T>30$~GeV &&\\ \hline
Cut-3 & 100~GeV $<H_T<$ 200~GeV  & \vline& $H_T >$ 100~GeV && \\ \hline
 
Cut-4           && & 96~GeV $< M_{(b\bar b)}< 136$~GeV  &&\\ \hline
Cut-5  & && $p_T^{b\bar b} >$ 75~GeV &&\\\hline\hline
\end{tabular}
\end{center}
\end{table}
\begin{table}
\caption{Number of signal for $\tilde{c}_{HW}$=0.1 and relevant backgrounds events after applied kinematic cuts used for the analysis at $\sqrt{s}$ = 380~GeV with $L_{int}=1$~ab$^{-1}$  for the three working point of b-tagging efficiency.\label{tab3}}
\begin{center}
\begin{tabular}{clccccccc}\hline\hline
Cuts &$b$-tagging eff. &$S+B_H$&$B_H$&$B_{ZZ}$&$B_{WW}$&$B_{tt}$&$B_{Z\nu\nu}$ \\ \hline
Cut-0 & - &69134&51743&288932&3059160&314891&298485 \\ \hline
          &50\% &10206&7664&10565&143&14861&13763  \\
Cut-1 & 70\%&20007&15036&20669&307&29643&27411\\
          & 90\%&33523&25086&39704&28764&62056 &49403 \\\hline
          &50\% &5376&3912&3004&82&6286&3963  \\
Cut-2 &70\%&10570&7637&5950&82&12271&7883\\
          & 90\%&17679&12669&10534&1657&24136 &14007 \\\hline
          &50\% &5130&3716&2603&61&2027&3360  \\
Cut-3 & 70\%&10074&7253&5115&82&3797&6613\\
          & 90\%&16858&12025&8891&1350&7351 &11736 \\\hline
     &50\% &4322&3162&131&-&776&144  \\
Cut-4& 70\%&8470&6171&225&-&1464&268\\
     & 90\%&14169&10207&432&123&2963 &522 \\ \hline
           &50\% &2971&1959&122&-&393&116  \\
Cut-5& 70\%&5815&3842&195&-&768&207\\
          & 90\%&9691&6352&381&20&1427 &417 \\ \hline\hline
\end{tabular}
\end{center}
\end{table}
Quantitatively, the final effect of all the cuts at the loose WP (90~\%) is approximately 14~\% and 12~\% for S+$B_{H}$($\tilde{c}_{HW}$=0.1) and $B_{H}$ while it is 0.1~\%, 0.4~\% and 0.1~\% for the $B_{ZZ}$, $B_{WW}$, $B_{tt}$, $B_{Z\nu\nu}$ backgrounds, respectively. 

\begin{figure}[hbt!]
\includegraphics[scale=0.18]{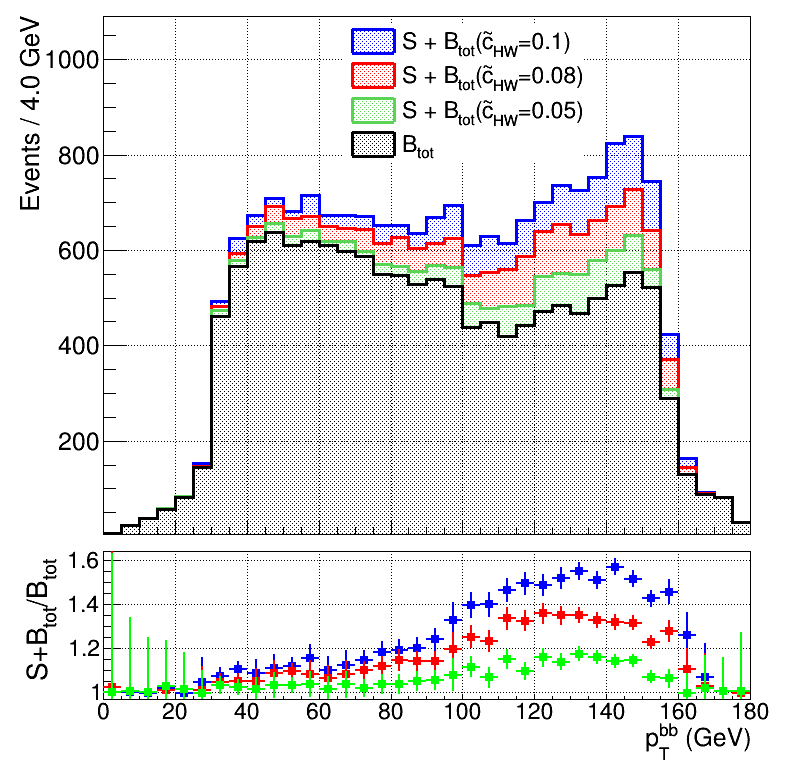} \includegraphics[scale=0.18]{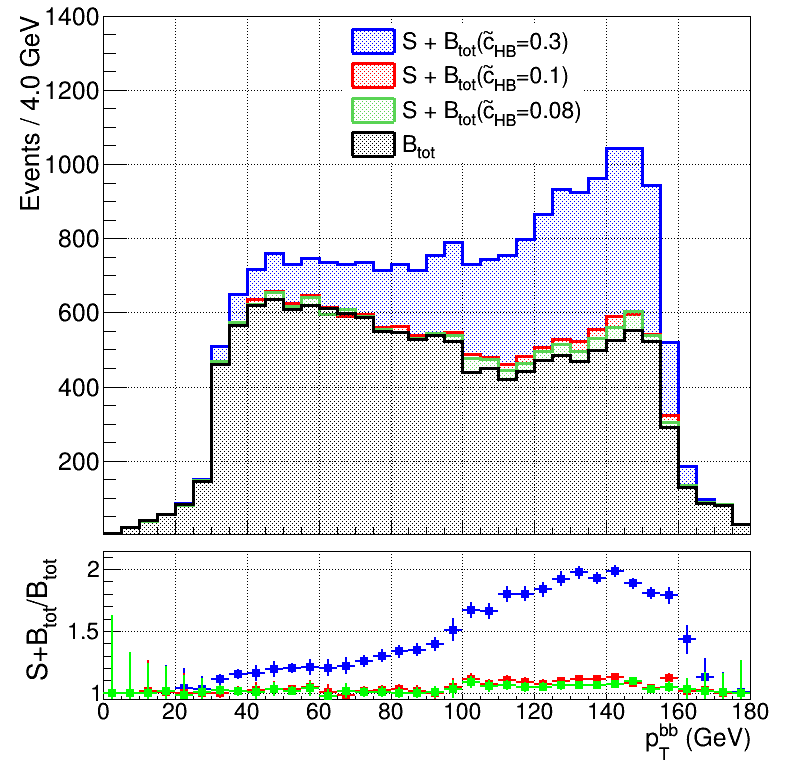} \includegraphics[scale=0.18]{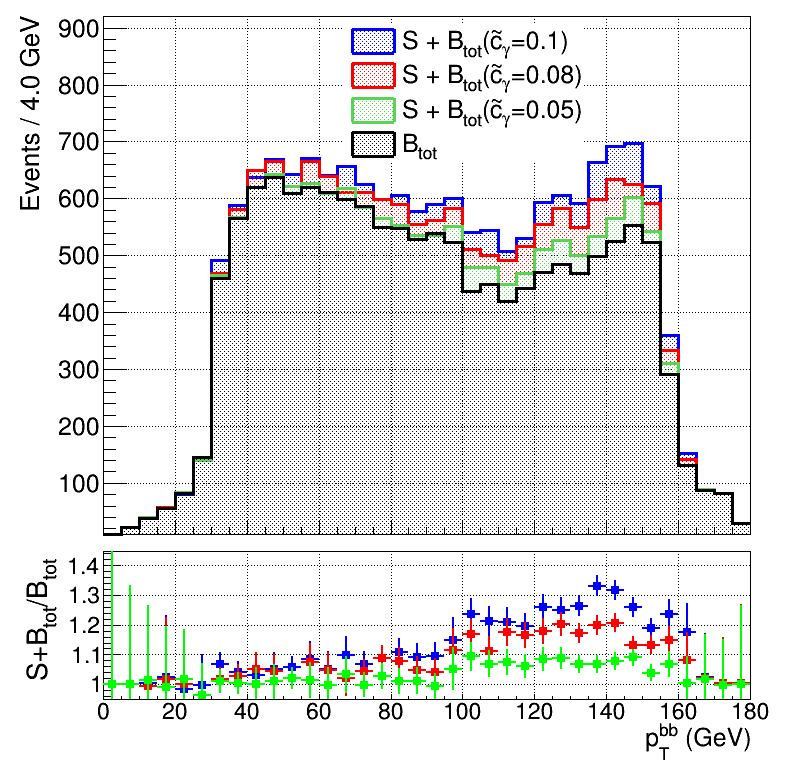} 
\caption{The transverse momentum distributions of the reconstructed Higgs boson of the signal for $\tilde{c}_{HW}$ = 0.05, 0.08 and 0.1;  $\tilde{c}_{HB} $ = 0.08, 0.1 and 0.3; $\tilde{c}_{\gamma}$ = 0.05, 0.08 and 0.1 couplings and relevant total SM background processes at $\sqrt s=$ 380~GeV. These distributions are normalized to $L_{int}=1.0$~ab$^{-1}$ \label{fig4}}
\end{figure}
After Cut-4, the transverse momentum distributions of the Higgs boson of the signal for $\tilde{c}_{HW}$ = 0.05, 0.08 and 0.1;  $\tilde{c}_{HB} $ = 0.08, 0.1 and 0.3; $\tilde{c}_{\gamma}$ = 0.05, 0.08 and 0.1 couplings and relevant total SM background processes ($B_{tot}$= $B_H$+$B_{ZZ}$+$B_{WW}$+$B_{tt}$+$B_{Z\nu\nu}$) are given in Fig.\ref{fig4}. The kinematic distributions shown in Fig.\ref{fig2}-\ref{fig4} are considered at the loose WP (90~\%) of b-tagging efficiency. All figures (Fig.\ref{fig2a}-\ref{fig4}) and number of events in Table \ref{tab3} are normalized to the cross section of each process times the integrated luminosity, $L_{int}$ = 1.0~ab$^{-1}$.
\subsection*{$\sqrt{s}$ = 1.5~TeV and 3~TeV}
 \begin{figure}[hbt!]
\includegraphics[scale=0.22]{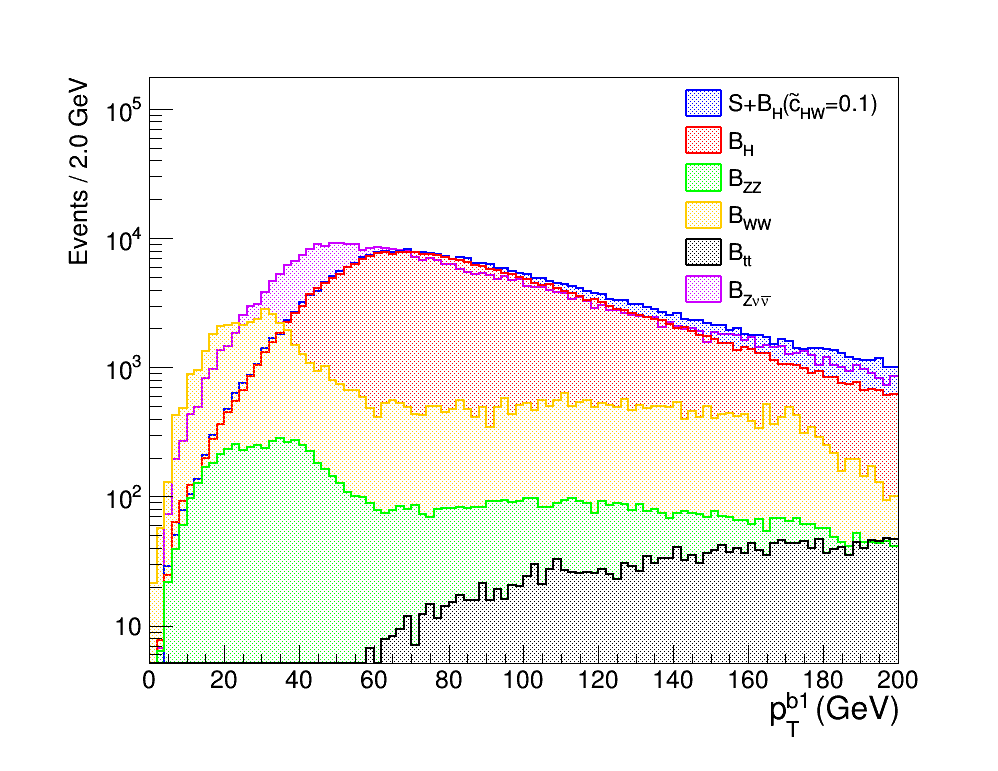} \includegraphics[scale=0.22]{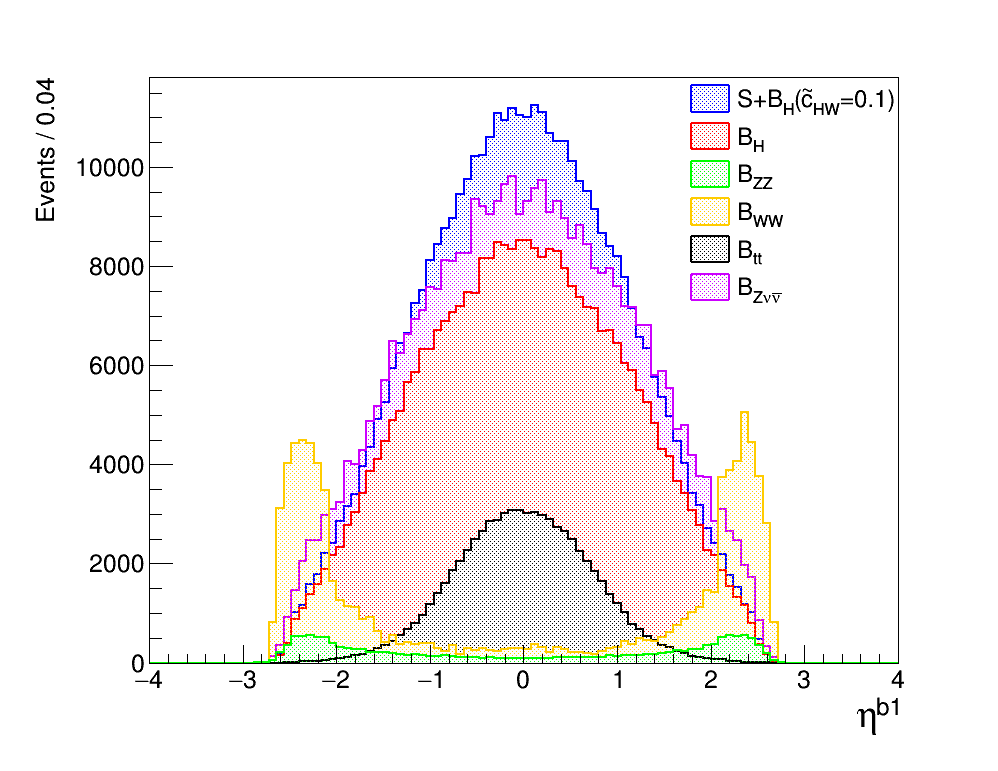}\\ 
\includegraphics[scale=0.22]{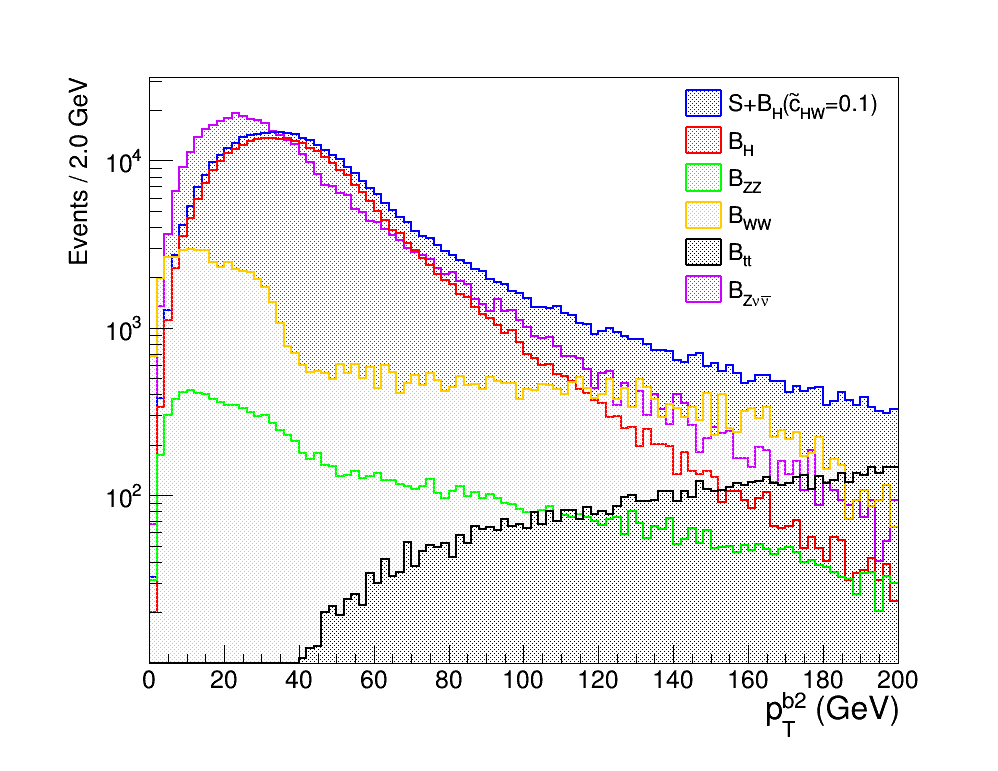} \includegraphics[scale=0.22]{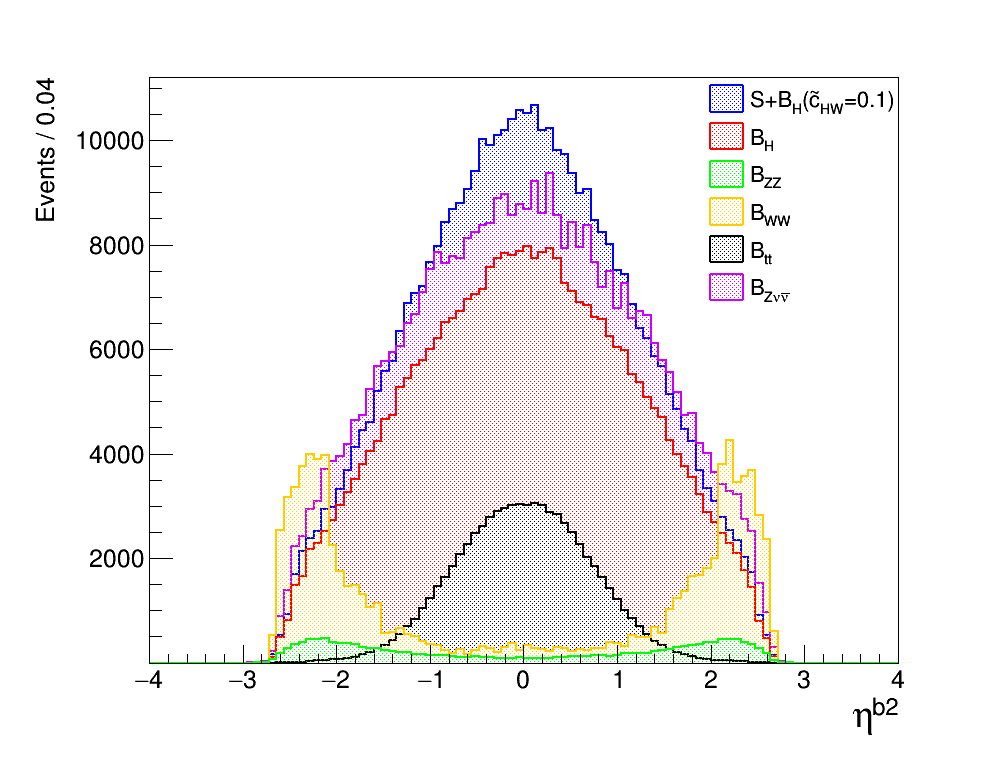} 
\caption{Normalized distributions of transverse momentum and pseudo-rapidity of tagged b-jets at $\sqrt s=$ 1.5~TeV; $b1$ (first row) and $b2$ (second row) for signal with $\tilde{c}_{HW} $=0.1, and relevant background processes. \label{fig5}}
\end{figure}
\begin{figure}[hbt!]
\includegraphics[scale=0.22]{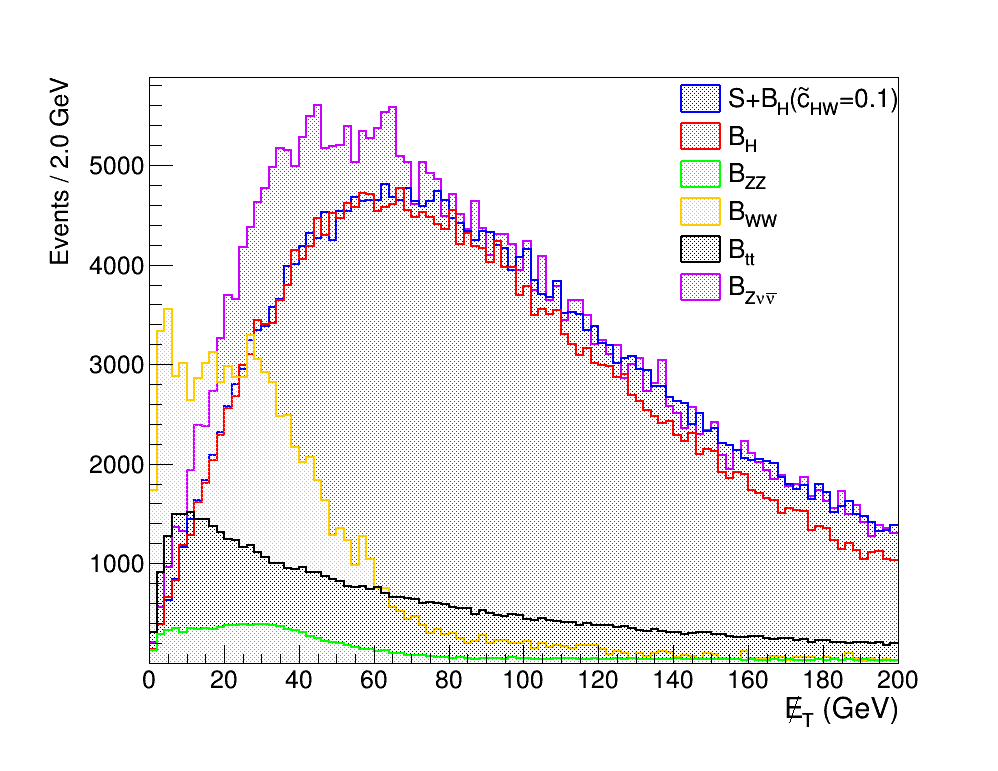} \includegraphics[scale=0.22]{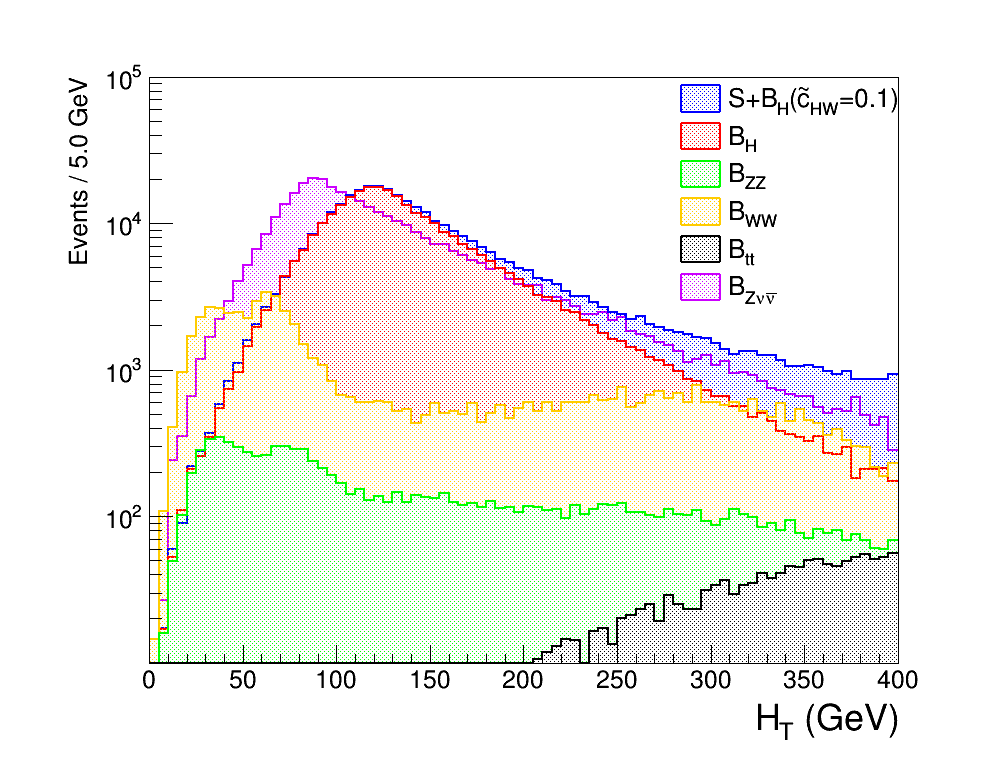} 
\caption{ Normalized distributions of Missing Energy Transverse  (${\not}E_T$), scalar sum of the transverse energy ($H_T$) at $\sqrt s=$ 1.5~TeV for signal with $\tilde{c}_{HW} $=0.1, and relevant background processes. \label{fig6}}
\end{figure}

 The analysis of $\sqrt{s}=$ 380~GeV is repeated for a 1.5 and 3 TeV center of mass energy of CLIC with $L_{int}=$ 2.5~ab$^{-1}$ and $L_{int}=$ 5.0~ab$^{-1}$, respectively. For the signal ($\tilde{c}_{HW}$= 0.1) and all relevant background processes taking the loose b-tagging working point at $\sqrt{s}=$ 1.5~TeV, the distributions of transverse momentum and pseudo-rapidity of $b1$ and $b2$ are shown in Fig.\ref{fig5} while the missing energy transverse (${\not\mathrel{E}}_T$) and scalar sum of the transverse energy ($H_T$) are given in Fig.\ref{fig6}. Both of these figures are normalized to the cross section of each process times the integrated luminosity, $L_{int}=$ 2.5~ab$^{-1}$. We only modified Cut-3 to $H_T> 100$~GeV as shown in Table \ref{tab2} at $\sqrt{s}=$ 1.5~TeV. Since similar distributions to Fig.\ref{fig5} and \ref{fig6} have been observed at $\sqrt{s}=$ 3~TeV, we implemented the same cuts used in the $\sqrt{s}=$ 1.5~TeV analysis.

\begin{figure}[hbt!]
\includegraphics[scale=0.18]{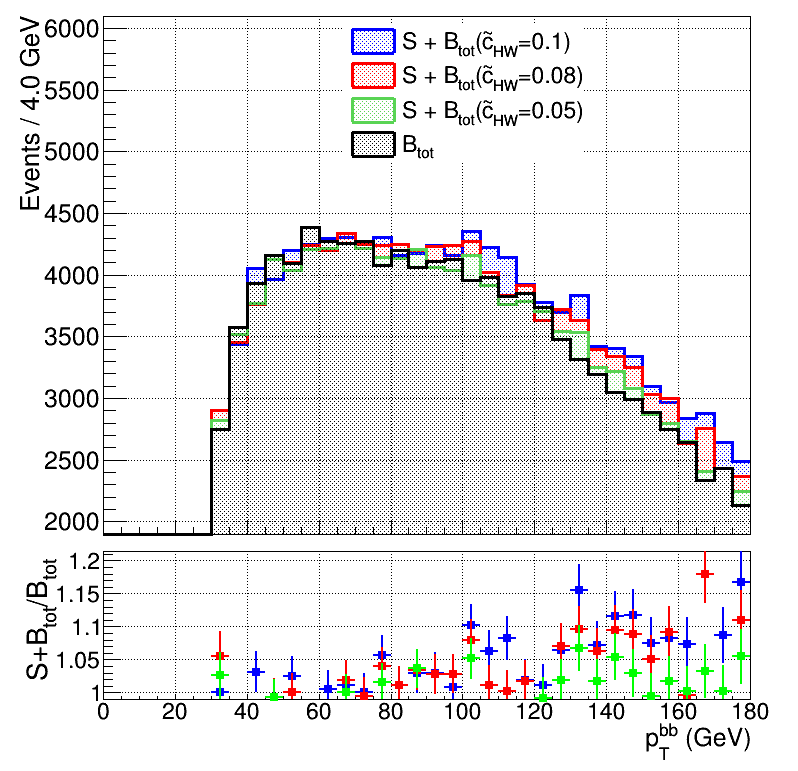} \includegraphics[scale=0.18]{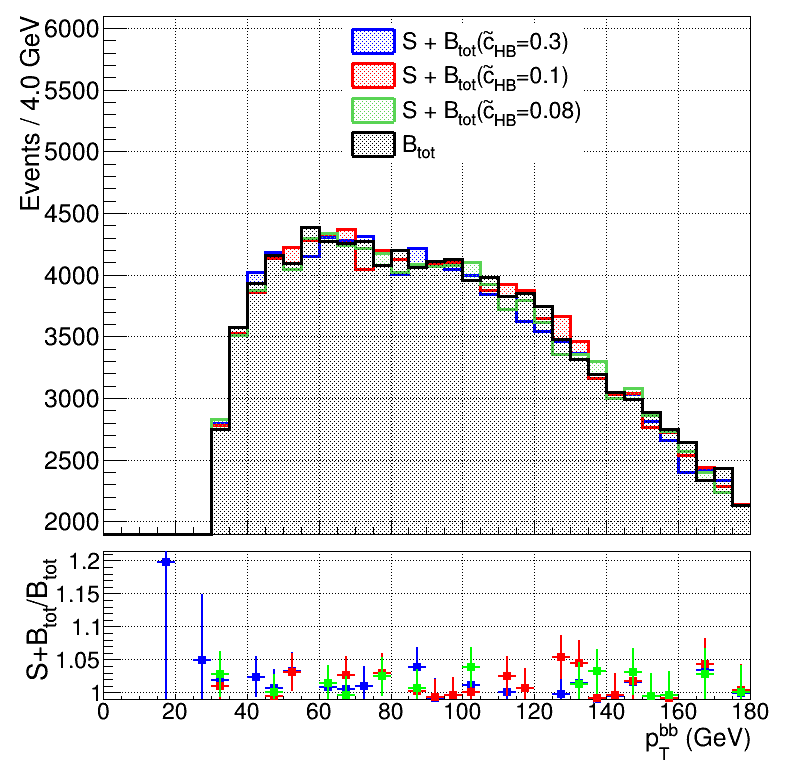} \includegraphics[scale=0.18]{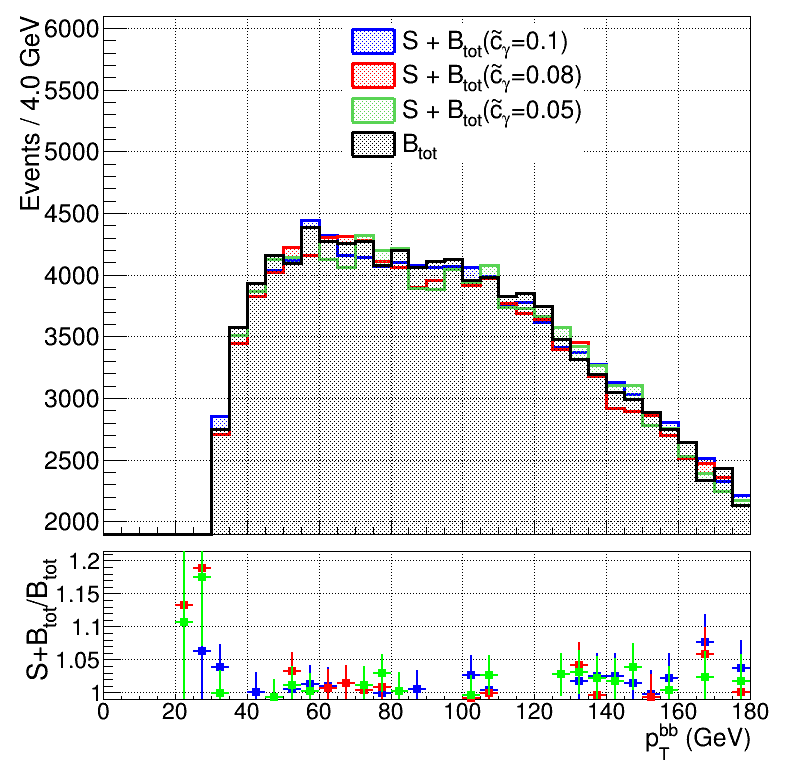} 
\caption{ The transverse momentum distributions of the reconstructed Higgs boson of the signal for $\tilde{c}_{HW}$ = 0.05, 0.08 and 0.1;  $\tilde{c}_{HB} $ = 0.08, 0.1 and 0.3; $\tilde{c}_{\gamma}$ = 0.05, 0.08 and 0.1 couplings and relevant total SM background processes at $\sqrt s=$ 1.5~TeV. These distributions are normalized to $L_{int}=2.5$~ab$^{-1}$ \label{fig7}}
\end{figure}

\begin{figure}[hbt!]
\includegraphics[scale=0.18]{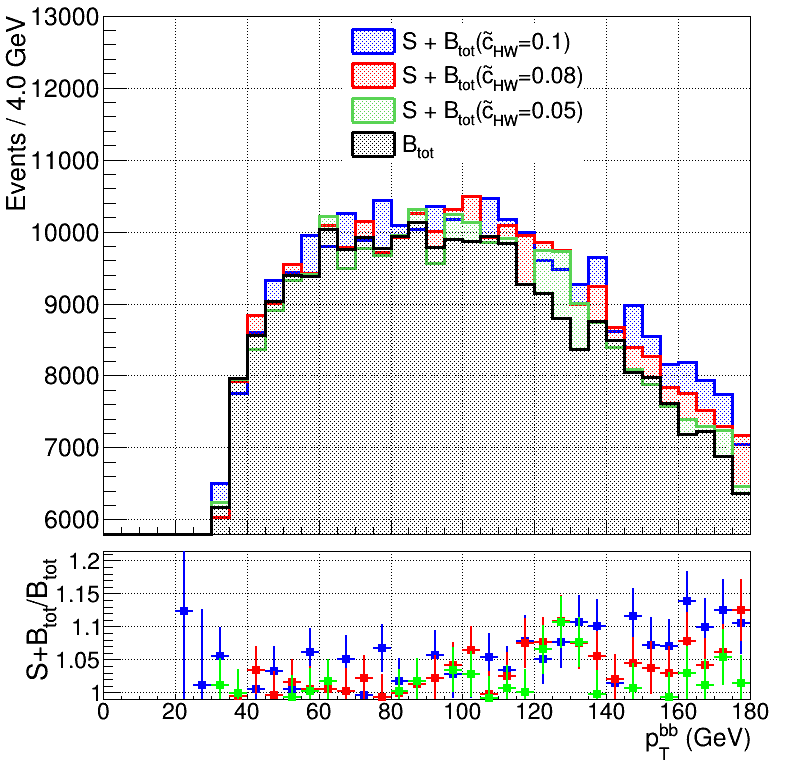} \includegraphics[scale=0.18]{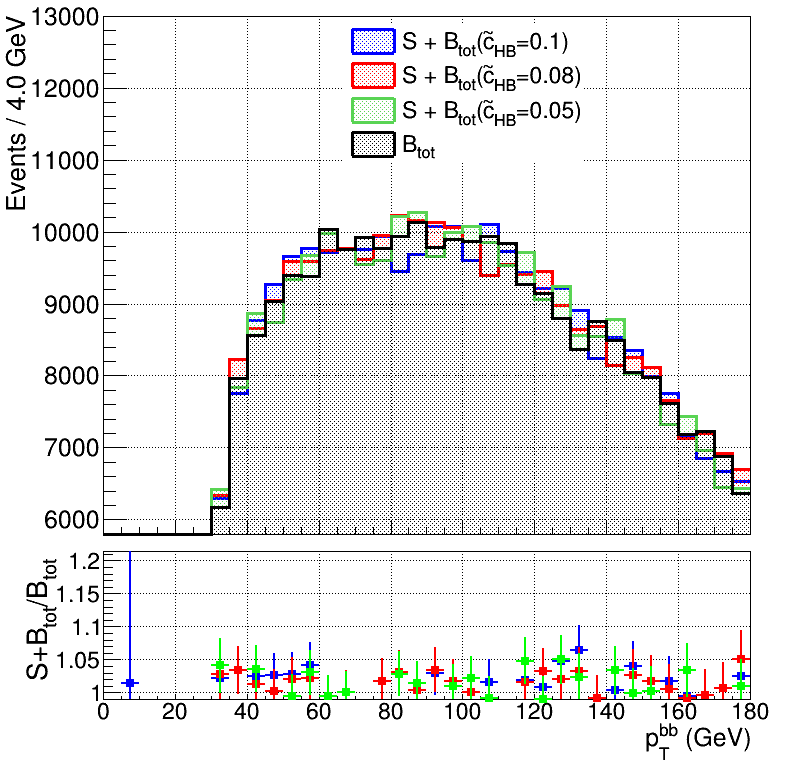} \includegraphics[scale=0.18]{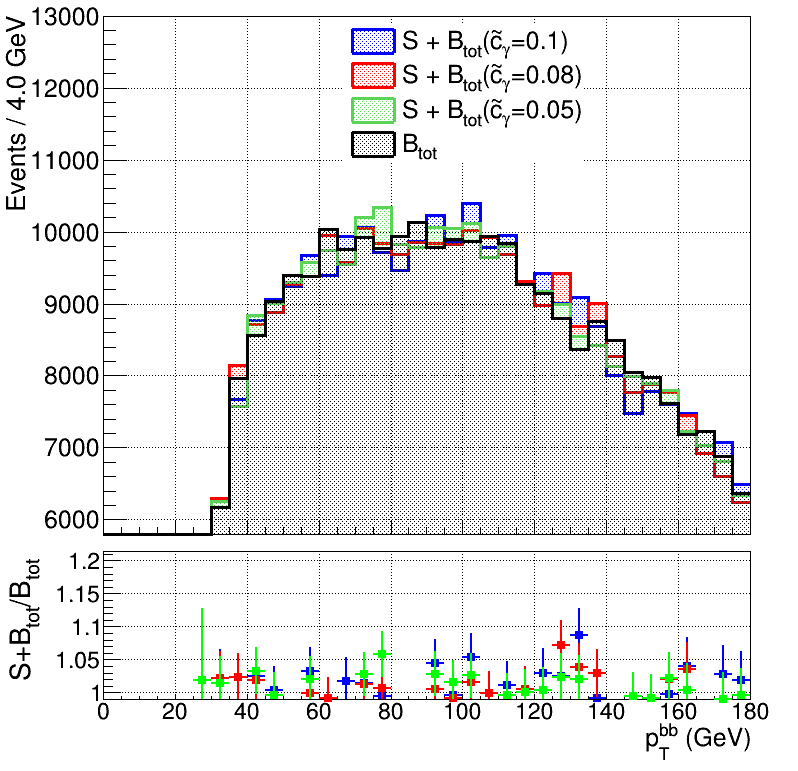} 
\caption{ The transverse momentum distributions of the reconstructed Higgs boson of the signal for $\tilde{c}_{HW}$ = 0.05, 0.08 and 0.1;  $\tilde{c}_{HB} $ = 0.08, 0.1 and 0.3; $\tilde{c}_{\gamma}$ = 0.05, 0.08 and 0.1 couplings and relevant total SM background processes at $\sqrt s=$ 3~TeV. These distributions are normalized to $L_{int}=5$~ab$^{-1}$ \label{fig8}}
\end{figure}

Finally, the transverse momentum distributions of the Higgs boson of the signal for $\tilde{c}_{HW}$ = 0.05, 0.08 and 0.1;  $\tilde{c}_{HB} $ = 0.08, 0.1 and 0.3; $\tilde{c}_{\gamma}$ = 0.05, 0.08 and 0.1 couplings (left to right) and relevant total SM background processes ($B_{tot}$) after Cut-4 are shown in Fig.\ref{fig7}-Fig.\ref{fig8} corresponding to 1.5 and 3~TeV center of mass energies, respectively. After applying the final cut, which requires the transverse momentum of $b\bar b$ system to be greater than 75 GeV, we obtained the normalized number of events for signals and relevant SM backgrounds. The total normalized number of events in the existence of effective couplings ($\tilde{c}_{HW} $=0.1, $\tilde{c}_{HB}$=0.3 and $\tilde{c}_{\gamma}$=0.3) and all relevant backgrounds are given in Table \ref{tab4}.

\begin{table}
\caption{Number of events for signal ($\tilde{c}_{HW} $=0.1, $\tilde{c}_{HB}$=0.3 and $\tilde{c}_{\gamma}$=0.3 couplings) and relevant backgrounds after Cut-4 for the analysis at $\sqrt{s}$ = 1.5  and 3~TeV with $L_{int}=2.5$ and 5.0~ab$^{-1}$ for the 90~\% working point of b-tagging efficiency. \label{tab4}}
\begin{center}
\begin{tabular}{lcccccccc}\hline\hline
$\sqrt s$ (TeV) &$S+B_H$ &$S+B_H$ &$S+B_H$ &$B_H$&$B_{ZZ}$&$B_{WW}$&$B_{tt}$&$B_{Z\nu\nu}$ \\ 
 &($\tilde{c}_{HW} $=0.1) &($\tilde{c}_{HB} $=0.3) &($\tilde{c}_{\gamma} $=0.3) &&&&& \\ \hline
   1.5       & 157328&152037&241080&100982&176 &-&69 &5943\\\hline
       3     & 535113&429096&623424&309473&115 &-&9 &20565\\\hline\hline

\end{tabular}
\end{center}
\end{table}


\section{Sensitivity of Higgs-Gauge Boson Couplings}
\begin{figure}[hbt!]
\begin{center}
\includegraphics[scale=0.45]{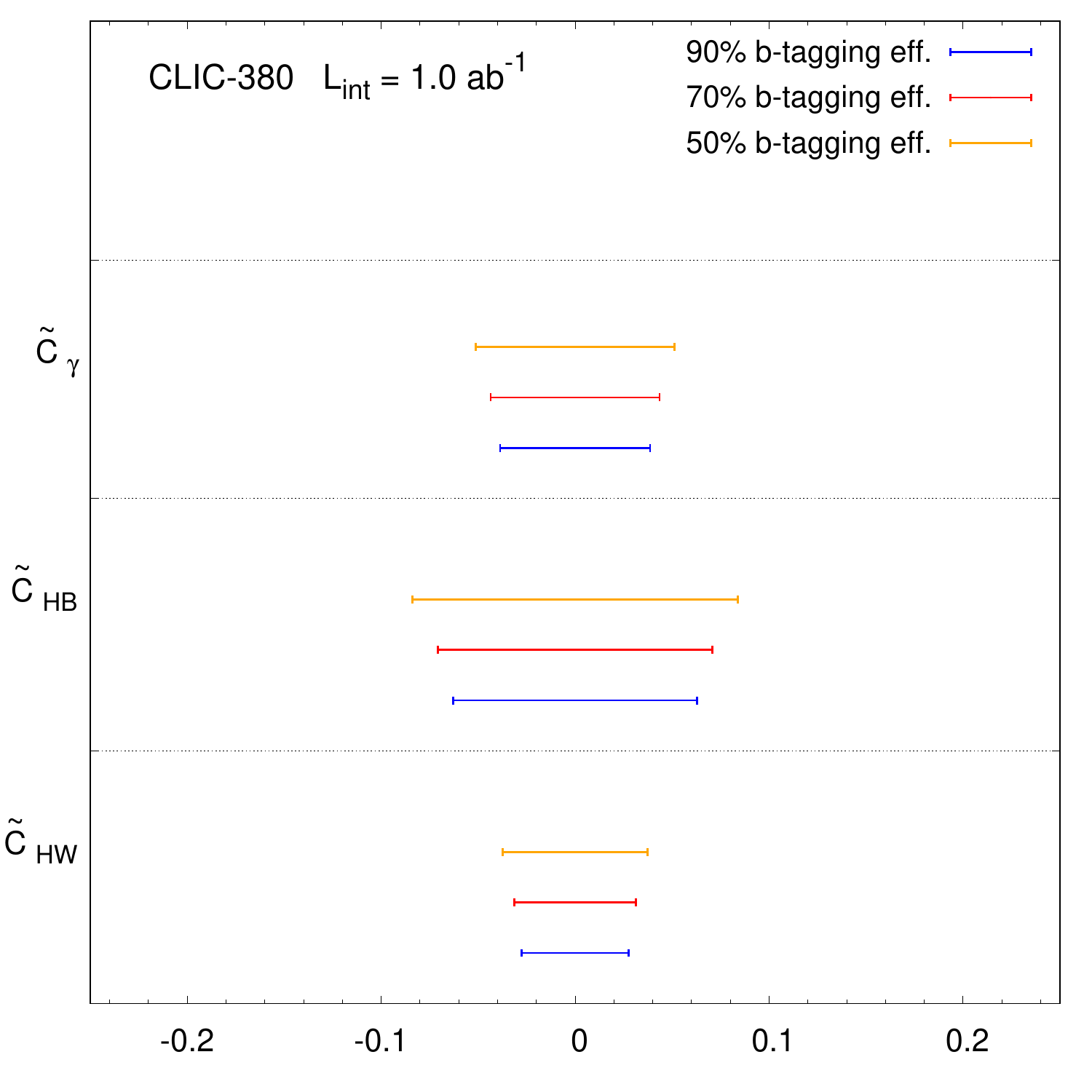} \includegraphics[scale=0.45]{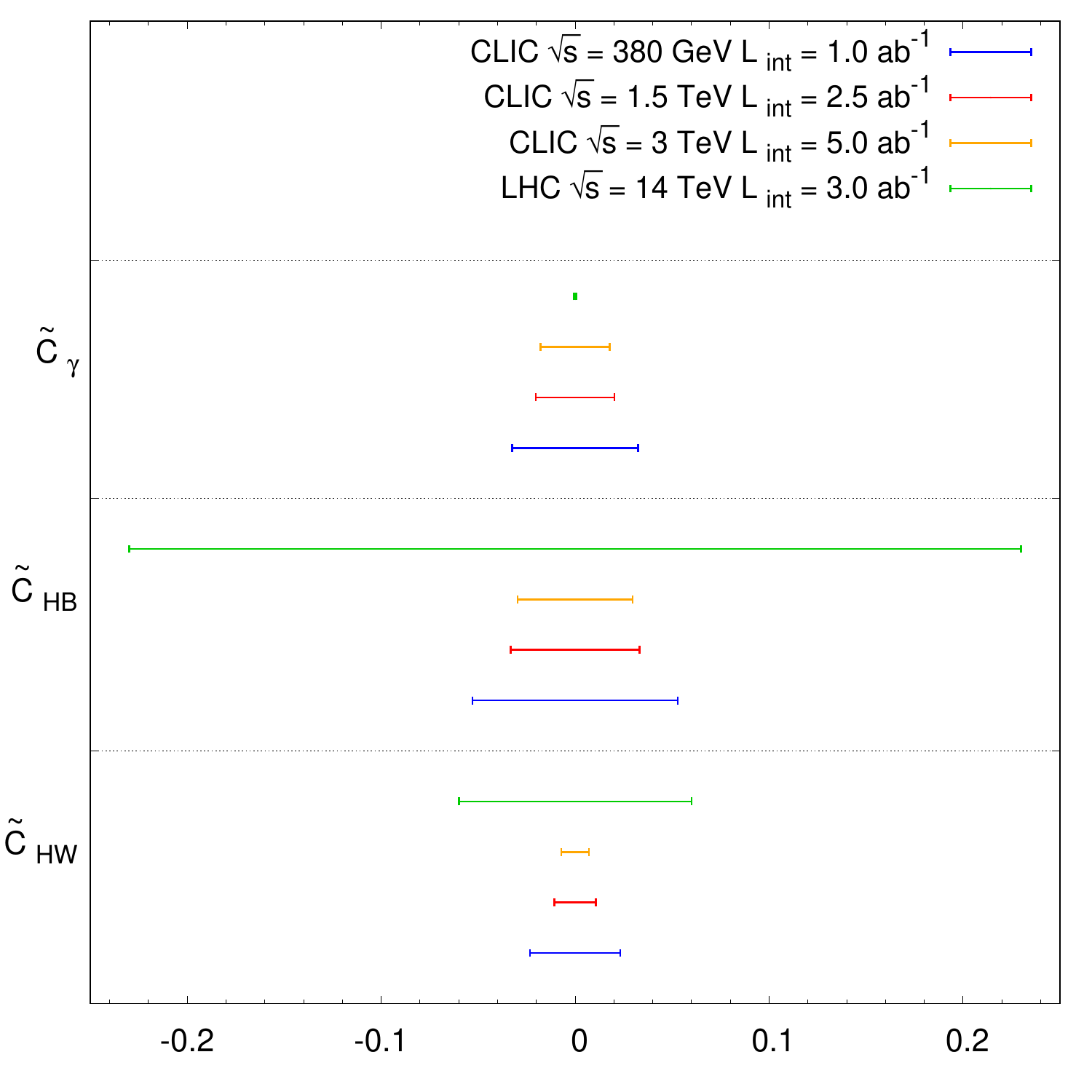} 
\caption{ Comparison of obtained 95~\% C.L. allowed range i) at the three working points of b-tagging efficiency for CLIC-380 with $L_{int}=1.0$~ab$^{-1}$ (on the left) ii) at 90~\% working point of b-tagging efficiency for all energy stages of CLIC compared with  HL-LHC projection limits at 14~TeV center of mass energy for the integrated luminosity of  3000~fb$^{-1}$ \cite{Ferreira:2016jea} (on the right), neglecting the effects of systematic uncertainties from both theoretical and experimental sources. \label{fig9}}
\end{center}
\end{figure}
The sensitivities to CP-violating dimension-6 Higgs couplings are obtained by a $\chi^2$ criterion method with systematic error, defined by 
\begin{eqnarray}\label{eq6}
\chi^{2} (\tilde c)=\sum_i^{n_{bins}}\left(\frac{N_{i}^{NP}(\tilde c)-N_{i}^{B_{tot}}}{N_{i}^{B_{tot}}\Delta_i}\right)^{2},
\end{eqnarray}
where $N^{NP}$ is the total number of events in the presence of effective couplings $(S)$ and total SM backgrounds $(B_{tot})$, $N^{B_{tot}}$ is the total number of events only coming from SM backgrounds, defined as $B_{tot}=B_H + B_{ZZ}+B_{WW}+B_{tt}+B_{Z\nu\nu}$, and $\Delta_i=\sqrt{\delta_{sys}^2+\frac{1}{N_i^B}}$ is the combined systematic ($\delta_{sys}$) and statistical error in each bin. In this study, we concentrate on obtaining $95~\%$ confidence level (CL) limits of the $\tilde c_{HW}$, $\tilde c_{HB}$, $\tilde c_{\gamma}$ couplings via the $e^+e^- \rightarrow \nu\bar{\nu} H$ signal process at CLIC with the center of mass energies at three stages $\sqrt{s}$= 380~GeV, 1.5~TeV, 3~TeV, and the integrated luminosities $L_{int}$= 1.0~ab$^{-1}$, 2.5~ab$^{-1}$, 5.0~ab$^{-1}$ respectively. Since we study the $H \to bb$ decay channel, b-tagging plays an important role in our analysis. To see the effect, we present the comparison of $b$-tagging efficiencies with three working points of $50~\%$, $70~\%$, $90~\%$ for the first stage center of mass energy of CLIC (CLIC-380) in the left panel of the Fig.\ref{fig9}. This figure emphasizes that the sensitivity of CLIC increases with the increase of $b$-tagging efficiencies, resulting in a better limit with the loose working point ($90~\%$ b-tagging efficiency). We measure the $H\nu\nu$ cross section in the channel $H\to b\bar b$ after b-tagging with statistical uncertainty of 1.67 \% in the first stage of CLIC for an integrated luminosity of 1 ab$^{-1}$ at $\sqrt s = 380$ GeV, assuming unpolarised beam and loose WP (90\%) of b-tagging efficiency. In the higher energy CLIC stages for integrated luminosity 2.5 ab$^{-1}$ at $\sqrt s = 1.5$ TeV  and 5 ab$^{-1}$ at $\sqrt s = 3$ TeV, the statistical uncertainties are 0.26 \% and 0.15 \%, respectively. In the right panel of Fig.\ref{fig9}, we plot obtained 95~\% C.L. limits at 90~\% working point of b-tagging efficiency for all three stages of CLIC and the recent High-Luminosity (HL-LHC) projections on these limits \cite{Ferreira:2016jea}. The HL-LHC projection limit on $\tilde c_{\gamma}$= $[-0.6\times10^{-3};0.6\times10^{-3}]$ is reported via $pp\to h\to\gamma\gamma$ process which is sensitive to this coupling.  However, we obtain better limits on $\tilde c_{HW}$, $\tilde c_{HB}$ than HL-LHC projection limits. At 3~TeV energy stage of CLIC, the sensitivities of $\tilde c_{HW}$ and $\tilde c_{HB}$ couplings are $[-7.0\times10^{-3};7.0\times10^{-3}]$ and $[-3.0\times10^{-2};3.0\times10^{-2}]$ with integrated luminosity of 5.0~ab$^{-1}$, respectively. Our limits on $\tilde c_{HW}$, $\tilde c_{HB}$ at $\sqrt s=3$~TeV with $L_{int}$= 5~ab$^{-1}$ are one order of magnitude better than HL-LHC projected limits and also better than observed current experimental limit on $\tilde c_{HW}$ (assuming $\tilde c_{HW}$=$\tilde c_{HB}$) measured  in the two-photon final state using 36.1~fb$^{-1}$ of proton-proton collisions at $\sqrt s = 13$~TeV by the ATLAS experiment at the Large Hadron Collider \cite{Aaboud:2018xdt}.
We also recomputed the bounds including systematic uncertainties at 95~\% C.L.. In the case of 0.3~\% systematic uncertainty from possible experimental sources as in Ref.\cite{Abramowicz:2016zbo}, the constraint on $\tilde{c}_{HW}$ and $\tilde{c}_{HB}$ at the highest energy stage of CLIC with $L_{int}$= 5.0~ab$^{-1}$ are $[-9.97\times10^{-3};9.97\times10^{-3}]$ and $[-4.18\times10^{-2};4.18\times10^{-2}]$. These bounds are lower than the experimental current limits and HL-LHC projected limits even at the first stage of CLIC. We should note that we neglect the theoretical uncertainties and only show the potential sensitivity of experimental reach. However, including theoretical systematic uncertainty is very likely to worsen our results.

The validity of the EFT can be tested with the relation between the new physics scale and the Wilson coefficients of the dimension-six operators as follows
 \begin{eqnarray}
 \bar c\sim\frac{g_*^2v^2}{\Lambda^2}
 \end{eqnarray}
where $g_*$ is the coupling constant of the heavy degrees of freedom with the SM particles. An upper bound on the new physics scale using $g_*=4\pi$ and obtained limits on $\tilde{c}_{HW}$ and $\tilde{c}_{HB}$ are 36.94 TeV and 17.84 TeV, respectively. This upper bounds are within the range of EFT.
\section{Conclusions}
  
  For a better understanding of the new physics beyond the SM in the Higgs sector, among the proposed future colliders, CLIC is an attractive option that has a clean environment. In this paper, we have emphasized the effects of CP-violating dimension-6 operators defined by an SM EFT Lagrangian approach via the $e^+e^- \rightarrow \nu\bar{\nu} H$ process for three energy stages of CLIC ($\sqrt{s}=$ 380~GeV, 1.5~TeV, 3~TeV)  and integrated luminosities ($L_{int}$= 1.0~ab$^{-1}$, 2.5~ab$^{-1}$, 5.0~ab$^{-1}$). We have presented the kinematical distributions of signal and relevant backgrounds; transverse momentum and rapidity of b-tagged quarks, missing energy transverse, scalar sum of the transverse energy and the invariant mass and transverse momentum of the Higgs boson, reconstructed from a pair of b-quarks (with 90~\% b-tagging efficiency). In order to obtain limits on the CP-violating dimension-6 couplings at each energy stage of CLIC, we focused on the transverse momentum of the reconstructed Higgs boson at three working points of b-tagging efficiency considering realistic detector effects with tuned CLIC detector cards designed for each center of mass energy stage in a cut-based analysis. The $e^+e^-\to\nu \nu H$ process is more sensitive to $\tilde{c}_{HW}$ and $\tilde{c}_{HB}$ couplings than the other CP-violating dimension-six couplings at three energy stages of CLIC. The obtained sensitivity of couplings at 95~\% C.L. of the $\tilde{c}_{HW}$ and $\tilde{c}_{HB}$ in all energy stages of CLIC are better than both HL-LHC projected and observed current experimental limits. As a conclusion, CLIC with three energy stages will offer advantages to probing the couplings of Higgs with SM particles that appear in the new physics beyond the SM scenarios.

\section*{Acknowledgments}
This work was supported by the Scientific and Technological Research Council of Turkey Turkish (TUBITAK), Grand No: 118F333. The authors would like to thank to CLICdp group for the discussions, especially to Philipp G. Roloff for valuable suggestions in the CLICdp Working Group analysis meeting. The authors would also like to thank to L. Linssen for encouraging us to involve in CLICdp collaboration. 
\printbibliography[title=References]
\end{document}